\documentclass[12pt,preprint]{aastex}

\shorttitle{\ion{Ne}{2} emission fro YSOs}
\shortauthors{Sacco et al.}
\usepackage{natbib}
\begin{document}


\title{High resolution spectroscopy of \ion{Ne}{2} emission from young stellar objects\thanks{Based 
on observations collected at the European Organisation for Astronomical
Research in the Southern Hemisphere, Chile, with the mid-infrared spectrograph and imager VISIR (Prog. ID 083.C-0883, 084.C-0211 and 085.C-0860)} \\}


\author{G.G. Sacco\altaffilmark{1,2}, E. Flaccomio\altaffilmark{3}, I. Pascucci\altaffilmark{4}, F. Lahuis \altaffilmark{5},
B. Ercolano\altaffilmark{6}, J.~H. Kastner\altaffilmark{1},  G. Micela\altaffilmark{3}, B. Stelzer\altaffilmark{3}, and M. Sterzik\altaffilmark{7}}


\altaffiltext{1}{Chester F. Carlson Center for Imaging Science, Rochester Institute of Technology,
    Rochester, NY 14623, USA}
\altaffiltext{2}{Present address: INAF-Osservatorio Astrofisico di Arcetri, Largo E. Fermi, 5, Firenze, 50125, Italy; gsacco@arcetri.inaf.it}
\altaffiltext{3}{INAF-Osservatorio Astronomico di Palermo, Palermo, 90143, Italy}
\altaffiltext{4}{Department of Planetary Sciences, University of Arizona, Tucson, AZ 85721, USA}
\altaffiltext{5}{SRON Netherlands Institute for Space Research, PO Box 800, 9700 AV Groningen, The Netherlands} 
\altaffiltext{6}{Ludwig-Maximilians-Universitaet University Observatory Munich D-81679, Muenchen, Germany }
\altaffiltext{7}{European Southern Observatory, Casilla 19001, Santiago 19, Chile}


\begin{abstract}

Constraining the spatial and thermal structure of the gaseous component of circumstellar 
disks is crucial to understand star and planet formation.
Models predict that the [\ion{Ne}{2}] line at 12.81 $\mu$m detected in young stellar objects
with Spitzer traces disk gas and its response to high energy radiation, but such [\ion{Ne}{2}] emission may also originate
in shocks within powerful outflows.
To distinguish between these potential origins for mid-infrared [\ion{Ne}{2}] emission
and to constrain disk models,
we observed 32 young stellar objects using the high resolution (R$\sim$30000) 
mid-infrared spectrograph VISIR at the VLT. We detected the 12.81 $\mu$m [\ion{Ne}{2}] line in 
12 objects, tripling the number of detections of this line in young stellar objects with high 
spatial and spectral resolution spectrographs.
We obtain the following main results: a) In Class I objects the [\ion{Ne}{2}] emission observed from Spitzer is mainly due to
gas at a distance of more than 20-40 AU from the star, where neon is, most likely, ionized by shocks due to protostellar 
outflows. b) In transition and pre-transition disks, most of the emission is confined to 
the inner disk, within 20-40 AU from the central star. c) Detailed analysis of line profiles 
indicates that, in transition and pre-transition disks, the line is slightly blue-shifted (2-12 $\rm km~s^{-1}$) 
with respect to the stellar velocity, and the line width is directly correlated with the disk inclination, as expected if 
the emission is due to a disk wind. d) Models of EUV/X-ray irradiated disks reproduce well the observed relation between the 
line width and the disk inclination, but underestimate the blue-shift of the line. 

\end{abstract}



\section{Introduction}

Disks orbiting young stars (1-10 Myr) are sources of accreting material
and sites of nascent planetary systems. Observations that can establish the physical
conditions and evolution of the gaseous component within young circumstellar disks are 
essential if we are to understand the magnetospheric accretion processes that 
determines a star's zero age main sequence mass, protoplanetary disk structure and evolution, and the processes involved in 
planet formation.
Despite their relevance to all of these important open issues, 
the physical and chemical properties of gaseous
circumstellar disks are still poorly understood. Submillimeter 
observations of molecular lines provide insight into the
chemistry and the physics of the outer disk regions (r$>$100 AU)
(e.g. \citealt{Dutrey:1994, Dutrey:1997, Kastner:1997, Kastner:2008}), 
but the properties of the gas within 20-30 AU have been only recently 
investigated thanks to the mid-infrared IRS spectrograph on board 
the \textit{Spitzer} Space Telescope, e.g., by the discovery of emission lines from several 
species including organic molecules and water \citep{Carr:2008, Salyk:2008}.

One of the most interesting mid-infrared spectral features 
is the [\ion{Ne}{2}] line at 12.81 $\mu$m \citep{Pascucci:2007, 
Lahuis:2007, Espaillat:2007, Flaccomio:2009, Gudel:2010}.
The importance of the [\ion{Ne}{2}] line is twofold: it traces warm gas (T$\sim$5000 K) 
and, due to the high first ionization potential
of neon (21.56 eV), it is a good proxy for the influence of stellar
extreme ultraviolet (EUV) and X-ray radiation on the disk.
The presence of strong [\ion{Ne}{2}] emission at 12.81 $\mu$m from protoplanetary 
disks was first predicted by \cite{Glassgold:2007}, who suggested that neon
is ionized by K-shell absorption of X-ray photons at energy E$>$0.9 keV.
This hypothesis has been refined by sophisticated irradiated-disk models
\citep{Meijerink:2008, Ercolano:2008, Ercolano:2010}, which predict a correlation
between the stellar X-ray luminosity and the total [\ion{Ne}{2}] line luminosity. 
It has also been suggested that X-rays may 
determine disk dissipation by heating the circumstellar gas to temperatures T$\sim10^3-10^4$ K and 
triggering its photoevaporation \citep{Ercolano:2008, Ercolano:2010, Owen:2010}. 
Neon ionization and disk photoevaporation could also be induced by EUV photons
produced by accretion shocks \citep{Hollenbach:2009, Alexander:2008}. However, EUV photons 
are more easily absorbed by neutral hydrogen, so they are less likely than X-rays to reach and affect the circumstellar disk,
especially early in the Class II phase when inflow and outflow from the inner disk
are densest \citep{Ercolano:2009, Owen:2010}.

A detailed comparison between 
models and observations is required to clarify this complex theoretical scenario.
Several surveys of the [\ion{Ne}{2}] emission from young stellar objects (YSOs)
have been carried out in the last 5 years, using low spectral and spatial resolution data 
obtained with \textit{Spitzer}/IRS \citep{Pascucci:2007, Espaillat:2007, Lahuis:2007, 
Flaccomio:2009, Gudel:2010}.
A weak correlation between X-ray and [\ion{Ne}{2}] luminosity has been found by
\cite{Gudel:2010}, who investigated a sample consisting of 92 pre-main sequence stars.
Furthermore, [\ion{Ne}{2}] emission was found to be stronger in Class I YSOs and
in sources driving protostellar jets than in more evolved YSOs \citep{Flaccomio:2009, Gudel:2010}. 
The strong [\ion{Ne}{2}] emission observed in embedded YSOs driving jets --
orders of magnitude larger than predicted by models of emission from disks --
is unlikely produced by irradiated gas within 
the inner disk, but it may be produced either by shocks, in the circumstellar envelope 
or in protostellar jets \citep{Hollenbach:2009}, or
in the launching region of magnetically driven outflows 
irradiated by stellar X-ray emission \citep{Shang:2010}.

The \textit{Spitzer}/IRS data, with their limited spectral and spatial resolution, provide 
limited information on the velocity and the spatial structure of the emitting gas.
High spectral and spatial resolution observations of the [\ion{Ne}{2}] line
from the ground can be used to decouple disk and jet emission. 
For example, using VLT/VISIR, clear evidence of [\ion{Ne}{2}] emission from protostellar 
jets has been discovered in the triple system T Tau \citep{Van-Boekel:2009}. 
The observations show that the predominant [\ion{Ne}{2}] emission component  
is shifted to a velocity compatible with the jet motion and is extended ($\sim1.1^{\arcsec}$) 
along the jet direction. 
However, a disk origin for the [\ion{Ne}{2}] emission has been suggested 
by high spectral resolution observations of the nearby classical T Tauri stars 
(cTTSs) TW Hya \citep{Herczeg:2007}, AA Tau and GM Tau \citep{Najita:2009}.  
In all of these stars the line appeared broadened ($\Delta v=21-70~km~s^{-1}$) 
and centered near the stellar velocity. The absence 
of detectable blue-shift suggested that the emission is produced from a static disk atmosphere,
but given the low signal-to-noise ratio of this data, it was not possible to rule out the presence 
of emission from a slow ($v=1-10~km~s^{-1}$) photoevaporative wind. 

Evidence of [\ion{Ne}{2}] emission arising from both a protostellar jet and a 
photoevaporative disk wind has 
been found by \cite{Pascucci:2009}, who observed 
4 optically thick disk objects (Sz 73, Sz 102, HD 3700A, VW Cha)
and 3 transition disk objects (TW Hydrae, T Cha and CS Cha). Transition disk objects have 
no excess in the near infrared and a strong excess in the far infrared, which 
is interpreted as a signature of a few AU central gap within the disk 
(see \citealt{Williams:2011} and references therein). 
\cite{Pascucci:2009} detected [\ion{Ne}{2}] emission from all the transition disk objects,
finding the line slightly blue-shifted ($\rm \Delta v\sim 4-6~km~s^{-1}$) and broadened, 
with a line width correlated with inclination angle. These characteristics appear 
consistent with model profiles expected from a photoevaporative disk wind triggered by
EUV or X-ray stellar emission \citep{Alexander:2008, Ercolano:2010}, but a larger number of analogous
observations are required to confirm their results. They detected 
the emission in only one of the optically thick 
disk objects, Sz 73. Since the emission is strongly blue-shifted 
($\rm \Delta v=100~km~s^{-1}$) and broadened ($\rm \Delta v=60~km~s^{-1}$),
an origin from fast outflowing material is more likely than a disk origin.
Additional evidence for a photoevaporative wind comes from the asymmetric [\ion{Ne}{2}]
line profile detected toward the transition disk around TW Hydrae \citep{Pascucci:2011}.

With the aim of understanding the origin of the [\ion{Ne}{2}] emission
from stars at various evolutionary stages and with different circumstellar characteristics
and to test X-ray and EUV irradiated disk models with observations,
we performed a high resolution  
spectroscopic survey of the [\ion{Ne}{2}] emission from 32 T Tauri stars and YSOs
spanning a range of spectral energy distributions (SED), with the VISIR spectrograph at the VLT. 
Here, we present the results of this survey, including 
a comparison with \textit{Spitzer}/IRS spectra, some of which are published here for the first
time.
In Sect. \ref{sec:obs}, we  describe the observed sample, the 
observations and the data reduction reduction procedure; in Sect. \ref{sec:results}
we report the results of analysis of the VISIR and the \textit{Spitzer} data;
in Sect. 4, we discuss our results and compare them with the predictions of irradiated disk models; 
and in Sect. 5, we draw our conclusions.

\section{Observations and data reduction \label{sec:obs}}

We observed 32 T Tauri stars and YSOs (2 resolved binaries IRS 43 and CrA IRS5 are counted as 4), belonging to several 
star forming regions and young stellar associations. 
In subsection \ref{sect:sample}, we describe the properties of the observed stars
and the procedures used to observe them; in subsection \ref{sect:data_red}, we describe 
the data reduction process.

\subsection{Target sample and observations \label{sect:sample}}

All the stars in our sample, with the exception of IRAS 08267-3336, are closer than 150 pc.  We include stars from both
very young star forming regions such as $\rho$ Oph and Corona Australis (1-3 Myr), which are characterized by 
a large number of deeply embedded protostellar objects, and older stellar associations like $\epsilon$ Cha and $\beta$ Pictoris
(6-12 Myr), that are mainly composed of apparently diskless (Class III) stars. We consider different regions and associations to 
include in our sample YSOs of different infrared Classes.
Furthermore, we gave higher priority to YSOs with \textit{Spitzer} detections of the
[\ion{Ne}{2}] emission at 12.81 $\mu$m. 
Two objects (T Cha and Sz 73) -- whose [\ion{Ne}{2}] line was already detected with VLT/VISIR 
by \cite{Pascucci:2009} -- have been observed again to investigate the variability of
the emission.

The full list of the observed targets is reported in Table \ref{obs-target}, where we list
object coordinates from the 2MASS catalog \citep{Skrutskie:2006}, parent region, distance 
(based on the parent region and data reported in the literature), heliocentric radial velocities, infrared Class,
disk inclination, X-ray luminosity in the band 0.3-10.0 keV\footnote{Some published L$_{X}$ values are for slightly 
different bands and were corrected to account for the band differences.} and mass accretion rate. All data in 
Table \ref{obs-target} are retrieved from the literature.
Radial velocities for some of the stars in the $\rho$ Oph and Corona Australis star forming regions were not available in the
literature, so these radial velocities have been estimated from the velocity of the parent cloud or nearby stars 
(see Table \ref{obs-target} for further details). Disk inclinations reported in Table \ref{obs-target} are derived by different techniques. Specifically, 
in 8 YSOs, disk inclinations are derived from interferometric submillimeter observations or scattered-light imaging of the disk  
(with estimated uncertainties 5-10 degrees). 
For RU Lupi we assumed the disk inclination to be equal to the inclination of the stellar rotation axes  (error $\sim$10 degrees), and 
for T Cha and RX J1615.3-3255 we report the inclination derived from a fit of the spectral energy distribution (SED).
Errors on the inclination derived from the SED are probably much higher. However, in the case of T Cha (inclination angle 75 degrees) other observational 
signatures -- e.g. a highly variable optical extinction -- demonstrate the presence of a circumstellar disk seen nearly edge-on.
For all the remaining targets, no information on the disk inclinations are available in the literature.

The infrared Class of our targets, reported in Table \ref{obs-target}, was derived from 
analyses reported in the literature (references are reported in the last column of Table \ref{obs-target}) 
of the SED in the near-infrared and mid-infrared bands and, in three cases (R CrA-7B, CrA IRS 5 and T CrA), of
mid-infrared color-magnitude diagrams.
Specifically, we classified, as Class I and Class II, sources with SED slopes in the ranges $\alpha>0$ 
and -2 $<\alpha\leq$0, respectively. To first approximation, the slope of the SED indicates the presence 
of a massive protostellar envelope (in Class I) or an optically thick disk (in Class II).
However, recent detailed studies of the dust emission in the mid-infrared and submillimeter band
(see \citealt{Williams:2011} for a review) suggest that within the Class II are included stars
with an optically thick disk and stars harbouring a more evolved
disk, characterized by the presence of an inner gap. 
As pointed out by \cite{Pascucci:2009}, the origin of the [\ion{Ne}{2}] emission in the latter group
of stars is much different than in the former, therefore, following the definitions mostly used 
in the literature, we define two more Classes of objects: the transition disks and the pre-transition disks. 
Specifically, circumstellar disks with signatures of a large (few AU) inner hole 
-- i.e. no excess in the near infrared and large excess in the mid infrared --- are classified as
transition objects (Tr), while circumstellar disks, with a SED indicating the presence of a gap, 
which separates the inner disk from the outer disk are classified as pre-transition disks 
(pTr, see \citealt{Espaillat:2007a} for a detailed discussion about the difference between transition
and pre-transition disks).

On the basis of these criteria, our sample is composed of 9 Class I, 13 Class II, 6 transition disks,
and 4 pre-transition disks. However, the information that we found in the literature for our sample
were not homogeneous so, we cannot exclude that with a more complete and homogeneous set of data some of the
Class II objects may be classified as transition or pre-transition disks.

It is worth noting that a large gap in the dusty component of the disk does not necessarily imply the presence of a gap in
its gaseous component, where the [\ion{Ne}{2}] emission is produced.
For example, in our sample T Cha, V4046 Sgr and Hen 3-600 are transition disk objects
but they are also actively accreting, as strong optical emission lines demonstrate \citep{Schisano:2009, Curran:2011}.  
The presence of these accretion signatures indicates that the hole within the dusty component of the disk is filled with gas.

VISIR observations have been carried out during three different ESO observing periods, P83, P84 and P85. 
We used VISIR \citep{Lagage:2004} in high resolution, long slit, spectroscopic
mode. The echelle grating is centered on the [\ion{Ne}{2}] fine structure line at 12.81355 $\mu$m
\citep{Yamada:1985} and covers the spectral range between 12.79 and 12.83 $\mu$m. 
We set the slit width at 0.4$^{\arcsec}$
($\sim$40 AU at a distance of $\sim$100 pc), which coincides with the spatial
resolution of the instrument when the visual seeing is better than 0.6-0.7$^{\arcsec}$ \citep{Smette:2007}, and 
a spectral resolution R$\sim$30,000 ($\rm \sim10~km~s^{-1}$ in velocity scale).
The spectral and spatial resolutions of our observations are, therefore, 50 and 10 times
higher than [\ion{Ne}{2}] observations carried out using IRS spectrograph on board \textit{Spitzer}, respectively 
(R$\sim$600 and slit width $\sim4.7^{\arcsec}$). 
For flux calibration and telluric correction, we observed a standard star immediately before and/or after
the observation of each target at airmass as close as possible to that of the target. The observation of stars
requiring long exposure times have been split in more than one segment, alternating an observation of the target 
with an observation of the standard star. 
Table \ref{Obs-log} reports date, time, airmass and exposure time of all the observed stars
and their associated standard stars. 
During the first run, we also observed the asteroid Kalliope to obtain a pure sky absorption spectrum,
and Titan to verify the instrument resolution.  Our Titan spectrum has been analyzed
by \cite{Pascucci:2011}, who measured the FWHMs and the centroid position of unresolved lines
of $\rm C_2H_2$ and $\rm C_2H_6$. The measured FWHMs span a very narrow range with a mean value of
10 $\rm km~s^{-1}$, as expected considering the instrument resolution, while the uncertainties on
the line centroids is $\sim$1 $\rm km~s^{-1}$. However, by tests on a set of standard stars and from 
TW Hydrae data, \cite{Pascucci:2011} found that the absolute accuracy of the line centroid is 2 $\rm km~s^{-1}$
due to uncertainties in placing the source in the center of the slit.

We also compiled fluxes in the [\ion{Ne}{2}] lines measured from \textit{Spitzer}/IRS \citep{Werner:2004,Houck:2004}
from the literature, and when these were not available, we 
downloaded, reduced and analyzed data from the \textit{Spitzer} archive.
\textit{Spitzer} [\ion{Ne}{2}] fluxes are reported in Table \ref{tab:Results}, together with the results obtained 
from the VISIR observations, for a direct comparison.

\subsection{Data reduction \label{sect:data_red}}

We reduced all VLT/VISIR data using the VISIR pipeline
Vers. 3.2.2 \citep{Lundin:2008} in conjunction with our IDL scripts.
VISIR data are a collection of data cubes, each corresponding to a nod position.
Every data cube contains 2n+1 planes, where n is the number of the chopping cycles. 
Odd planes contain the images from the on source chopping cycle ($A_i$), while the 
even planes contain the average of the differences between the current and all the previous 
on source ($A_i$) and off source ($B_i$) chopping cycle images. 
Therefore, the last plane contains the average of all the $A_i-B_i$ images.

The first part of the data reduction is executed by the pipeline. Specifically, all 
the data cubes are coupled in pairs, composed of two complementary nod positions.
The last planes of each pair are summed and divided by 
the total integration time.  At this point, because we used standard observing mode, 
in which the chopping and nodding directions and throws are the same, each nodded image 
contains a double positive beam in the center and a negative beam on each side. 
To form the final science frame, all the 
nodded images are added, after bad pixel cleaning and correction of 
the image distortion. 
In a small number of cases, we excluded from the final sum some pairs of nod images taken 
when the background emission was particularly strong and variable, e.g. due to the presence of thin clouds. 
The standard deviation of the noise in the resulting image is calculated from the dispersion
of pixel values, after the rejection of outliers with an iterative sigma-clipping.
Pixels with values less than 3 times the standard deviation of the noise are classified as noisy.

\begin{figure*}
\includegraphics[angle=0, width=16 cm]{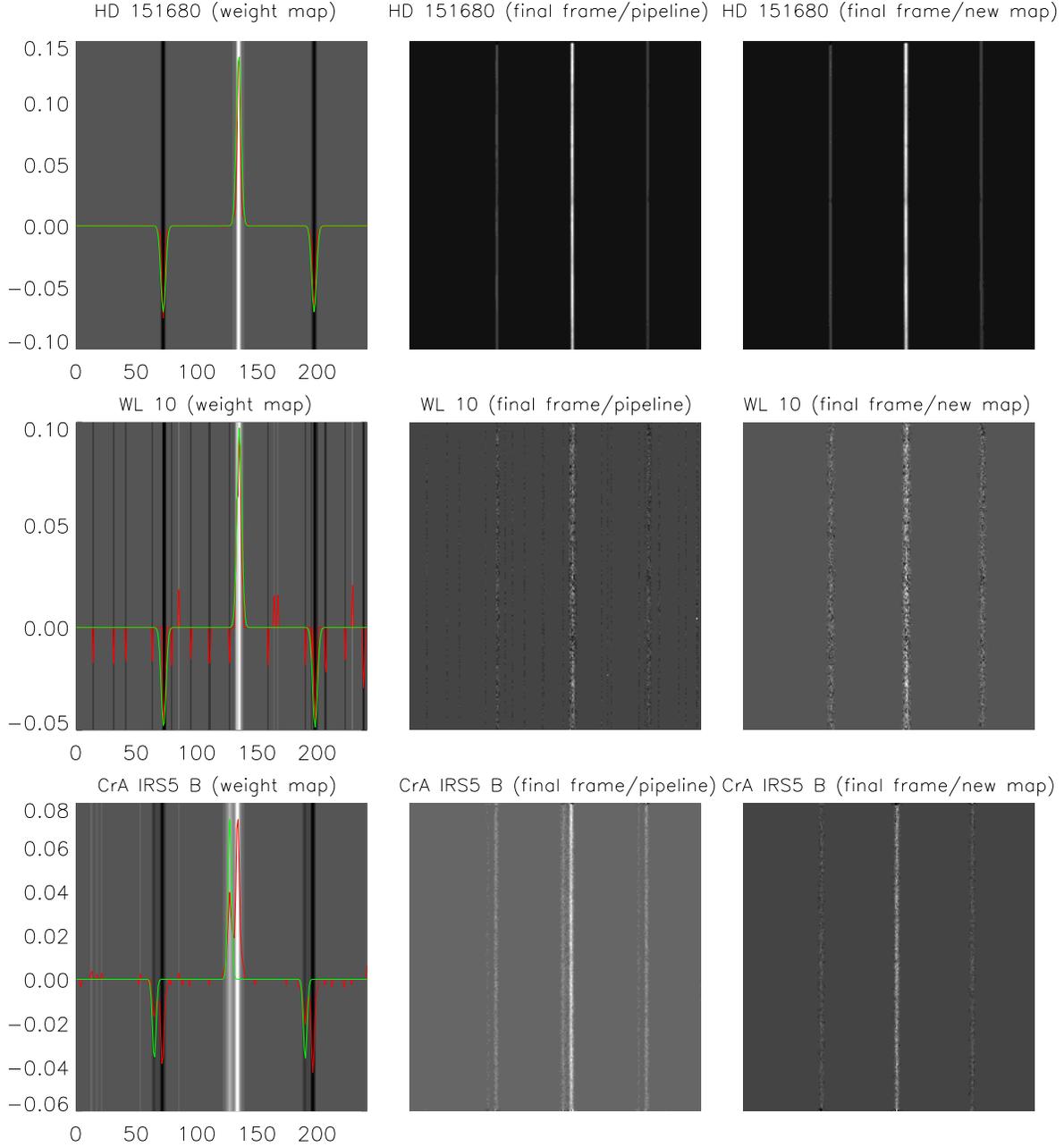}
\caption{Examples of reduction of VISIR spectral image data. The left panels show weight maps calculated by the pipeline, while the central and the right panels
show the final 2D spectrum obtained by the pipeline and by an alternative data reduction using our own weight map, respectively. 
From top to bottom we show the frames for a very bright star (the standard HD 151680),
a very faint star (WL 10) and a binary (CrA IRS 5), respectively.
The red and the green lines overplotted on the left panels are the profiles along the spatial direction 
of the weight maps calculated from the pipeline (red line) and defined by us (green line). The two profiles
are normalized to the same maximum value. On the x and y axes, we report the number of pixel and 
the intensity of the profile of the weight, respectively. 
For the binary, we show only the weight map for secondary component. 
\label{fig:data_red}}
\end{figure*}

The spectrum is extracted by the pipeline using the method developed by \cite{Horne:1986}. 
Specifically, a weight map is generated by collapsing the
spectrum along the dispersion direction, normalizing the absolute flux of the
one-dimensional image to 1 and setting to zero noisy pixels.
The resulting profile is then replicated along the dispersion direction to cover the full frame.
The final spectrum is obtained by multiplying the science frame for the weight map
and summing along the slit direction. Pixels classified as noisy are used to calculate errors 
(see Fig. \ref{fig:data_red}).  

After several tests, we realized that this method for spectrum extraction fails in two situations:  
a) when the signal-to-noise ratio 
is very low and the residual noise is comparable to the signal from the star,
because the weight map, and therefore the extracted spectrum, is significantly
contaminated by noise (see the the middle panels of Fig. \ref{fig:data_red});
b) when we observe a spatially resolved binary with the two components aligned along the slit, 
because by multiplying the 
science frame for the weight map and summing along the slit direction
we obtain an integrated spectrum of both components, losing information on the
emission of each component.
To avoid these problems, we extracted the spectra of faint sources and resolved binary systems by defining our own weight map. 
The weight map for faint sources is the sum of three gaussians, one centered on the
peaks of the positive beam and two centered on the negative beams, with half and opposite amplitudes with respect to the first one 
(see Fig. \ref{fig:data_red}). 
To extract separately the spectra of each component of the spatially resolved binaries,
we defined different maps for each component, using the same method described above (see bottom panels of Fig. \ref{fig:data_red})
To check this procedure, we applied this method to the bright stars of our sample; obtaining 
the same results as with the standard procedure (see top panels of Fig. \ref{fig:data_red}).

The wavelength calibration is done by 
cross-correlating an infrared background spectrum with a synthetic model spectrum of the atmosphere.
The infrared background spectrum is extracted from the input files before correcting for chopping and nodding.
The whole extraction procedure described above is performed in the same way for both the science target and the 
associated standard stars. 
Finally, the flux calibrated spectrum is obtained by multiplying the observed spectrum of the star by a 
spectral response function. The response function is calculated by dividing
the observed spectrum of a standard star by the flux model (at very low resolution and not including photospheric 
absorption lines) retrieved from the VISIR standard stars catalog. 
Photospheric absorption lines in the spectra of the standard stars have been eliminated by gaussian fitting.
This procedure allow us to correct both for sky transmission, which depends on the wavelength - because of 
telluric lines - and on the airmass, and for fringing, a flux modulation with the 
wavelength that is characteristic of VLT/VISIR spectra and that has been found to be constant over periods of 
several months \citep{Van-Boekel:2009}.
This procedure may not be very accurate if the signal-to-noise ratio of the standard is very low, if the airmass of the standard
is different from the airmass of the target, or if the photospheric lines of the standard are not well
subtracted, because they are blended with telluric lines.
To verify that none of these sources of error affect our results, we derive our calibrated
spectra using an alternative approach. Specifically, we remove the fringing and the telluric lines using a 
fringing model and a model of the atmosphere and, then, we used the standard stars only for the flux calibration. A more 
detailed discussion of this alternative method is reported in \cite{Pascucci:2011}.

The two methods for deriving the calibrated spectra give the same results for all the stars except 
for T CrA. In this case, the first procedure did not work very well because the airmass of the star and of
the standard are different and because the equivalent width of the [\ion{Ne}{2}] line is very low,
so systematic errors associated with flux calibration turn out to be larger than statistical errors. 
On the basis of these considerations, for T CrA we applied the second procedure for the flux calibration.

\textit{Spitzer}/IRS data for stars observed in the c2d survey were reduced using the c2d data reduction pipeline
(see \citealt{Lahuis:2007} and \citealt{Lahuis:2006} for further details).
Data for other stars were downloaded from the Spitzer Heritage Archive in the basic calibrated data (BCD) format processed 
with the S18.18.0 data pipeline. These data were analyzed using custom IDL code (including SSC software 
IRSCLEAN MASK) and the SMART (v.8.2.1.) program developed by the IRS Team at Cornell University 
(\citealp{higdon:2004}; \citealp{lebouteiller:2010}).
To further improve the flux calibration, we reduced 4 spectra of the standard star $\xi$ Draconis using the same procedure as
for the science targets and, then we recalibrated the observed flux (the median among the 4 spectra) by using the median of these
spectra and a model spectrum, that has been produced by MARCS code with the specific purpose of calibrating \textit{Spitzer} spectra
\citep{Decin:2004}.

\section{Results \label{sec:results}}

We observed 32 YSOs. Our targets belong to
the star forming regions Taurus, $\rho$ Oph, Corona Australis, Lupus and the Gum nebula and to the nearby associations 
TW Hya, $\epsilon$ Cha and $\beta$ Pic. 

\begin{figure*}
\includegraphics[angle=0, width=17 cm]{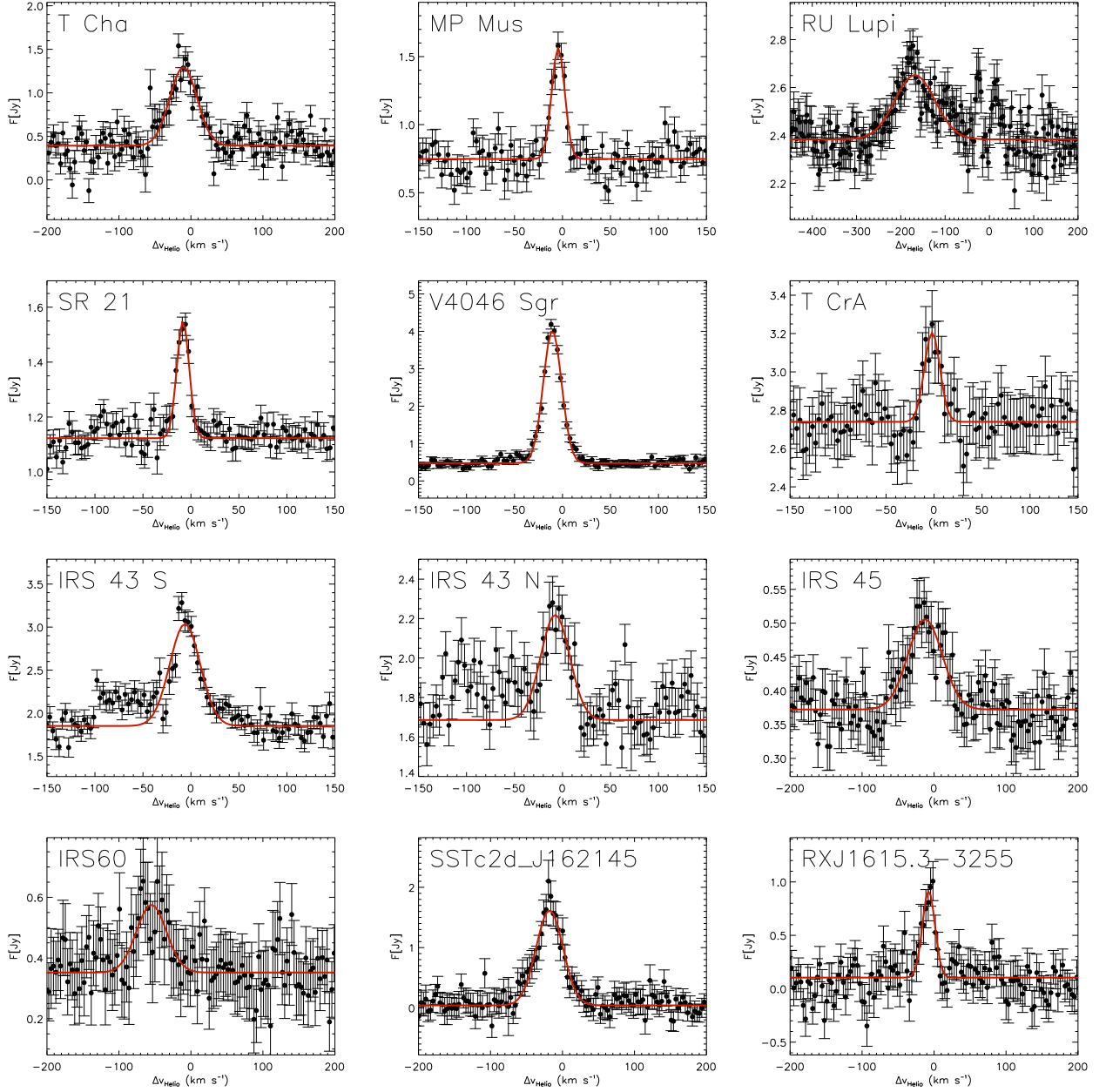}
\caption{Flux calibrated spectra of the 12 sources, for which we detected the 12.81 $\mu$m [\ion{Ne}{2}] emission line. The 
gaussian best fit of the data is shown by the red curves. The spectra are plotted as function of radial velocity
in the stellocentric reference system.
\label{fig:spectra}}
\end{figure*}

\begin{figure*}
\includegraphics[angle=0, width=15 cm]{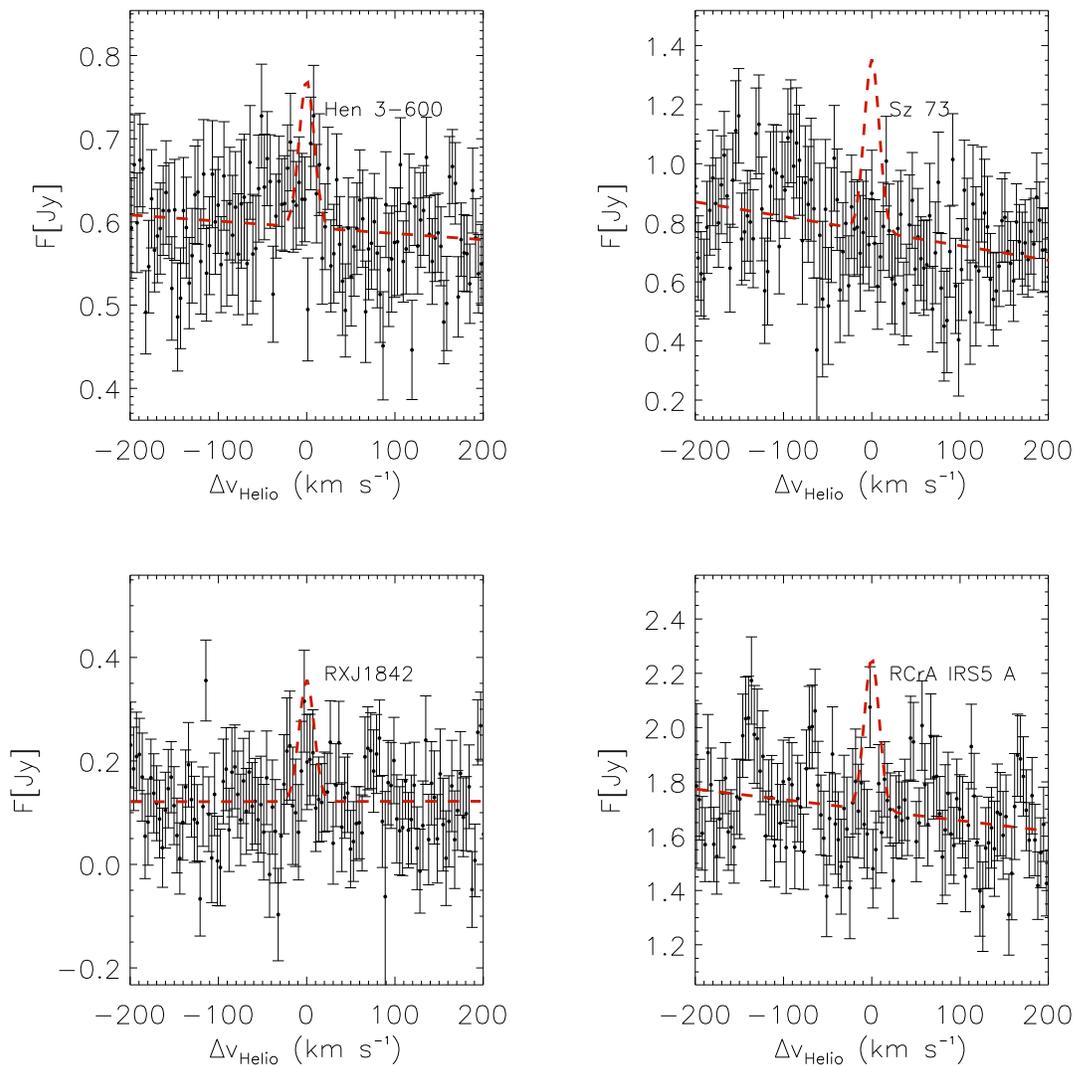}
\caption{Flux calibrated spectra of the 4 sources for which we did not detect the 12.81 $\mu$m [\ion{Ne}{2}] emission line. 
Overplotted red lines show gaussian profiles of a line with a total flux equal to the upper limit and a FWHM=20$\rm km~s^{-1}$. 
Radial velocities (x axes) are with respect to the observed star as in Fig. \ref{fig:spectra}.
\label{fig:no_det}}
\end{figure*}

Our results are summarized in Table \ref{tab:Results}. Fluxes, full widths at half maximum (FWHMs) and 
peak velocities (v$_{peak}$) of the emission lines are calculated by fitting the spectral lines with the sum of a gaussian 
and a constant continuum (see Fig. \ref{fig:spectra}).
Errors (1$\sigma$) in the total flux and FWHM are derived by a best fit procedure. Systematic errors associated with photometric calibration will be discussed later in
this section. 
Errors on line peak velocities are the sum, in quadrature, of the statistical errors associated with the fitting procedure,
the error on the stellar velocity reported in Table \ref{obs-target}, and an additional error of 2 $\rm km~s^{-1}$ 
that accounts for errors in positioning the star within the slit (see \citealt{Pascucci:2011} and Sect. \ref{sect:sample}). 
If in the observed wavelength range there are no significant emission features, we assume that the line has not been detected and 
we report the upper limit in Table \ref{tab:Results}.
The upper limit depends on the assumed line width and was computed as $F_{up}=5 \sigma_{con}\sqrt{FWHM \delta v}$,
where $\sigma_{con}$ is the standard deviation of the continuum emission, FWHM is the assumed full width at half maximum of the 
line and $\delta$v is the width of a velocity bin ($\rm \sim 3.3~km~s^{-1}$). 
Upper limits reported in Table \ref{tab:Results} are calculated for an assumed FWHM=20 $\rm km~s^{-1}$. 
Four spectra of stars, for which [\ion{Ne}{2}] was not detected are shown in Fig. \ref{fig:no_det} overplotted with a gaussian
with a total flux equal to the upper limit and a FWHM=20 $\rm km~s^{-1}$. We chose to plot and report the
upper limit for an assumed FWHM=20 $\rm km~s^{-1}$, because it is the typical FWHM of the observed emission line
produced by photoevaporative winds (see Sect. \ref{sect:inner_disk}).

Even if standard stars were observed just after or before the science targets and were chosen
as close as possible to the targets, a residual systematic error due to discrepancy between the airmass
of the standard and of the target could affect our results. 
\cite{Pascucci:2011} studied the systematic errors on VISIR observations of the [\ion{Ne}{2}] emission by
observing TW Hya several times during the same night. In their observations both flux and line peak velocity
depended on the airmass. Specifically, the flux decreased of up a factor 2 as the airmass 
increased from 1 to 1.4, while the peak velocity varied from -4 to -7 $\rm km~s^{-1}$. In contrast, \cite{Pascucci:2011} 
found that the observed FWHM of the line did not depend on the airmass.

All the detected lines are shown in Fig. \ref{fig:spectra} with best-fit gaussians overlaid on the data. 
All lines are well fitted by a gaussian, except
RU Lupi and IRS 43. Specifically, the emission of RU Lupi looks asymmetric, with
the line peak blue-shifted with respect to the line center; while in the spectra of 
both IRS 43 N and IRS 43 S we observe two features, a stronger one, which peaks
very close to the stellar radial velocity and is well fitted by a gaussian profile, 
and a weaker one, which is strongly blue-shifted ($\rm v_{peak}\sim-100~km~s^{-1}$)
and asymmetric. However, this strongly blue-shifted component is very weak and we are not sure if it is a real feature
or a statistical fluctuation. 

Two of our targets (Sz 73 and T Cha) have been previously observed by \cite{Pascucci:2009}, using VISIR.
To test our data reduction procedure and to compare our observations with the previous ones, we rereduced again the
data obtained by \cite{Pascucci:2009}, using our version of the pipeline. We obtained slightly higher fluxes in both cases;
these are the values reported in Table \ref{tab:Results}.
The new version of the pipeline performs a better subtraction of the sky that can influence the observed
flux of very faint targets like T Cha and Sz 73. However, these discrepancies are not significant, so they do
not influence the scientific results of either ours or previous analysis. 

The two datasets for T Cha and Sz 73 also offer an opportunity to investigate variability
of the [\ion{Ne}{2}] line. In the case of T Cha the line flux, shape and FWHM remained the
same. On the contrary, in Sz 73 we failed to detect the [\ion{Ne}{2}] from the outflow 
as previously reported in \cite{Pascucci:2009}. The second epoch data show upper limits
of a factor 1.6 lower than the epoch one data (see Table \ref{tab:Results}). 
Because emission from Sz 73 is due to a protostellar jet,
different slit orientations, different seeing or different positions of the star within the slit
could produce significant discrepancies in the measured line flux. We can rule out the first two hypotheses, because 
in the two observations the slit orientation on the plane of the sky was the same and the seeing was slightly better  
during the second observation when the line was not detected. Errors in centering the star within the slit or intrinsic 
variability could explain the discrepancy between the results of the two observations.

In Table \ref{tab:Results}, we also report line luminosities (or their upper limits), 
calculated from the distances reported in Table \ref{obs-target}. We also consider
the effect of the absorption from interstellar and circumstellar gas, which, at 12.81 $\mu$m, 
is not negligible for highly embedded sources, i.e when the absorption in the J band is A$_{J}>1$ mag
\citep{Flaccomio:2009}. 
Specifically, we obtained A$_{J}$ (see Table \ref{tab:Results}) from the literature and, 
when it was not available, from measurement of optical extinction, 
using the conversion law $\rm A_{J}=0.282 A_{V}$ \citep{Rieke:1985}, or 
from the absorption in the X-ray band ($\rm N_{H}$), using the conversion law $ \rm A_{J}=1.8 \times 10^{-22} N_{H}$ 
\citep{Vuong:2003}. 
The absorption A$_{J}$ is larger than 1 mag in 13 YSOs in  
$\rho$ Ophiuchi, 3 sources in the Corona Australis 
and, only marginally (A$_{J}=1.4$ mag) in IRAS 08267-3336. \cite{Chapman:2009} showed that
for highly absorbed stars in $\rho$ Oph and other star forming region an extinction law
with $R_{V}=5.5$ is appropriate, as it accounts for grain growth in the dense part of the cloud. 
Therefore, for sources with $\rm A_{J}>1$, we calculate the absorption at 12 $\mu$m, using the relation $\rm A_{12.81}=0.16 A_{J}$ 
from \cite{Weingartner:2001}, resulting in an absorption $\rm A_{12.81}$ ranging from 0.22 to 4.16 mag. For
all other sources, we used the standard $R_v=3.1$ extinction
law ($\rm A_{12.81}=0.097 A_{J}$ from \citealt{Weingartner:2001}). However, for these sources, the 
absorption has a negligible effect ($\rm A_{12.81}<0.1$ mag), independent of the extinction law used for the 
calculation.

In Table \ref{tab:Results}, we also report [\ion{Ne}{2}] fluxes measured from 
spectra taken by \textit{Spitzer}/IRS in the high resolution mode (R$\sim$600). 
We used values either reported from the literature or measured from 
archival data. For measuring line fluxes or their upper limits from \textit{Spitzer} data, we used
the same procedure as in \cite{Flaccomio:2009}. Specifically, we calculated the flux by integrating the
emission between 12.78 and 12.84 $\mu$m and subtracting the underlying continuum by fitting a first 
or a second order polynomial in two adjacent spectral intervals (12.72-12.78 and 12.84-12.91 $\mu$m). 
In two cases (SSTc2dJ162145.1 and RX J1615.3-3255) for which high resolution spectra were not available, 
we used the low resolution spectrum and different
wavelength ranges to estimate the line flux (12.7-12.9 $\mu$m) and the continuum   
(12.3-12.7 and 12.9-13.3 $\mu m$). Upper limits (3 $\sigma$) have been calculated, as previously 
discussed for VLT/VISIR data, from the rms on 
the continuum emission and assuming a fixed line width (0.023 $\mu$m).

\section{Discussion}

We detected the [\ion{Ne}{2}] line in 12 YSOs (3 Class I, 4 Class II,
4 transition disks and 1 pre-transition disks); 11 of those are new detections with VISIR.
We measured the FWHM and the velocity shift of all the detected lines. In Sect. \ref{sect:spitz_vs_visir},
we compare fluxes of the [\ion{Ne}{2}] line measured with VISIR and \textit{Spitzer} and we discuss
how discrepancies between observations with the two instruments can depend on the spatial location 
of the [\ion{Ne}{2}] emitting gas. In Sect. \ref{sect:inner_disk}, we compare all our results with other 
ancillary data (disk inclination, X-ray luminosity and mass accretion rate) and irradiated disk models
to better understand the physical mechanisms generating [\ion{Ne}{2}] emission from gas within the inner disk.

\subsection{Origin of the [\ion{Ne}{2}] emission \label{sect:spitz_vs_visir}} 

The aim of this work is to understand the origin of the [\ion{Ne}{2}] emission in YSOs
during the different phases of their evolution. Different emission mechanisms have been discussed in the literature.
The [\ion{Ne}{2}] emission could be a tracer of shocks produced by the interaction between 
a protostellar outflow and the interstellar medium \citep{Hollenbach:1989, Hollenbach:2009}. This mechanism has been
suggested to be at the origin of the emission observed from the triple system T Tau \citep{Van-Boekel:2009}. The main observational
evidence supporting this hypothesis is the spatial location of the emission: in fact, the emission is
extended (1.1$^{\arcsec}$) and located along the stellar outflow. Otherwise, the neon
could be ionized by high energy stellar emission. In this scenario the
[\ion{Ne}{2}] emission at 12.81 $\mu$m should be produced within the inner circumstellar disk, few AU from the
central star. A handful of observations of transition disks support this second scenario by showing that 
the emission is usually compact and at a velocity close to that of the radial velocity of the
central star \citep{Herczeg:2007, Najita:2009, Pascucci:2009, Pascucci:2011}.

\begin{figure*}
\includegraphics[angle=0, width=15 cm]{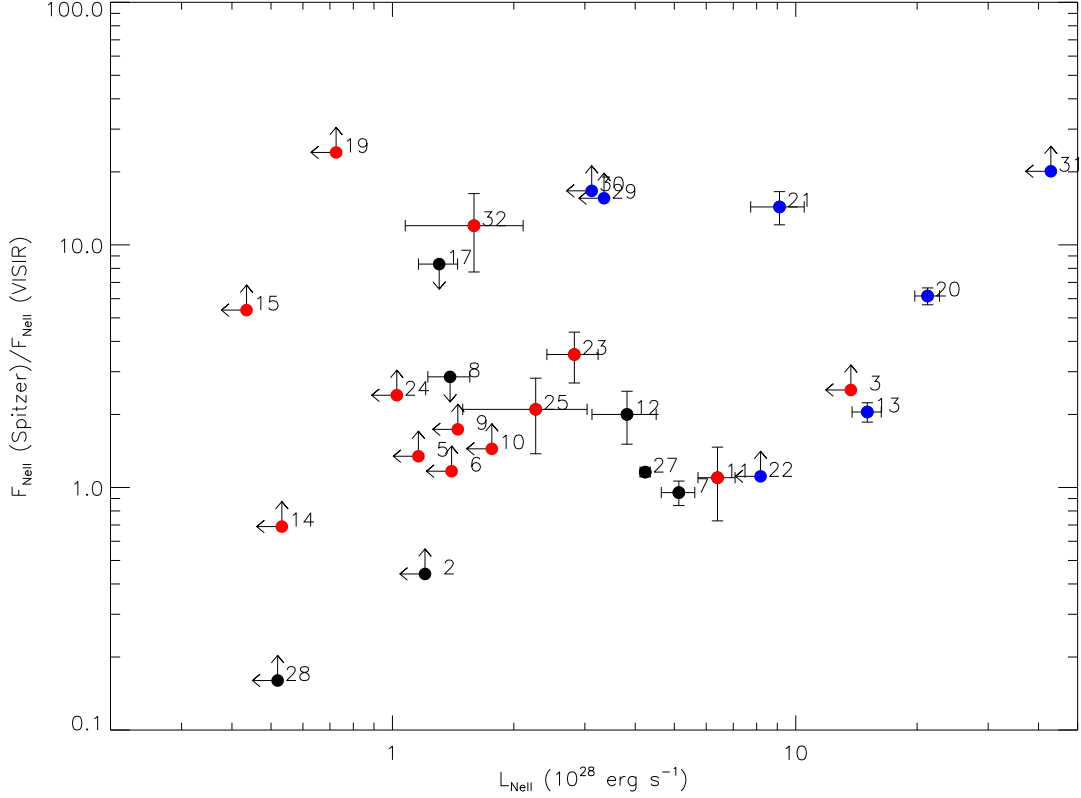}
\caption{Ratio between flux of the [Ne II] line as measured from \textit{Spitzer}/IRS and VLT/VISIR spectroscopic data as function
of the [Ne II] luminosity measured from the VISIR data. Blue and red dots indicate Class I and Class II YSOs, respectively, while transition and 
 pre-transitional disks are both indicated with black dots. For the two binary systems (IRS 43 and CrA IRS5) not resolved 
by Spitzer, but resolved by VISIR, we assumed that the [Ne II] emission observed by Spitzer was equally shared by the two 
components. Numbers are the IDs of the stars reported in Table \ref{tab:Results}.
\label{fig:visir_vs_spitzer}}
\end{figure*}

Information on the spatial distribution of the [\ion{Ne}{2}] emission can be obtained by comparing fluxes measured 
by \textit{Spitzer} (spatial resolution $\sim$4.7$^{\arcsec}$) and VISIR (spatial resolution $\sim$0.4$^{\arcsec}$). 
To illustrate this, we show in Figure \ref{fig:visir_vs_spitzer} the ratio between 
the absorbed flux observed using \textit{Spitzer} and VISIR as function of the [\ion{Ne}{2}] absorption-corrected luminosity measured from the
VISIR spectra for YSOs of various infrared Class. A systematic discrepancy is evident between the
[\ion{Ne}{2}] fluxes measured by \textit{Spitzer} and VISIR for Class I YSOs (blue dots); namely,
\textit{Spitzer} fluxes range from 2 to more than 20 times larger than [\ion{Ne}{2}] fluxes measured by VISIR,
with the exception of one object, which might thus 
be compatible with the other Class I stars.
This discrepancy strongly suggets that the [\ion{Ne}{2}] emission observed from \textit{Spitzer}
in most Class I objects is spatially extended, i.e. it is not produced within the inner disk (r$<$20-40 AU), 
but in the outer regions of the circumstellar envelope. Alternatively, the emission may span a very broad range 
of velocities, such that the upper limits reported in Table \ref{tab:Results} could be underestimated 
(up to a factor 5 if we consider a line marginally resolved from \textit{Spitzer}/IRS at high resolution mode). 
However, both scenarios would suggest that neon in class I objects is most likely ionized by shocks produced 
by protostellar outflows.
To better investigate the [\ion{Ne}{2}] emission from shocks located several AUs from the star, 
a different observational strategy is required. As previously done for T Tau by \cite{Van-Boekel:2009},
the position and the velocity structure of multiple knots composing a jet and emitting in the [\ion{Ne}{2}] 
can be studied by performing several long slit observations 
with different slit orientations to cover a large area around the target. Furthermore, 
to perform a complete study of the physical structure of protostellar jets, the
results of these observations can be compared with narrow band imaging and spectroastrometric
observations in other forbidden lines, like the [\ion{S}{2}] doublet at 6717 and 6731 \AA~or the 
[\ion{O}{1}] at 6300 \AA.

On the other hand, in YSOs which are classified as transition or pre-transition disks 
(Tr and pTr in Table \ref{obs-target} and black dots in 
Fig. \ref{fig:visir_vs_spitzer}), the ratio
between [\ion{Ne}{2}] fluxes measured by the two instruments is consistently less than 2, 
with the exception of two stars for which the
upper limits might be compatible with other transition and pre-transition disks. For the
3 objects detected with both instruments, the ratio is consistent with 1 at the 3 $\sigma$ confidence level.
We conclude that, in transition and pre-transition disks, the bulk of the [\ion{Ne}{2}] 
emission is produced within $\sim$20-40 AUs of the central star.
In the next section, we will discuss if this emission is due to a static disk atmosphere, 
a photoevaporative wind, or a magnetically accelerated outflow.

It is more complicated to interpret results for Class II YSOs, namely stars harbouring an optically thick disk.
We detected the [\ion{Ne}{2}] emission in 4 Class II YSOs. Since the ratio between Spitzer and VISIR fluxes 
ranges from $\sim$1 to $\sim$10, the spatially extended emission appears to be dominant, but significant 
emission from the inner disk may also be present.

\subsection{[\ion{Ne}{2}] emission from the inner region: static disk atmosphere, photoevaporative wind or irradiated protostellar jet 
\label{sect:inner_disk}}

As discussed in the previous section, the detection of spatially unresolved [\ion{Ne}{2}] emission in most of 
the transition and pre-transition disks demonstrates 
that the [\ion{Ne}{2}] line is produced within the inner disk (less than 20-40 AUs from the star) suggesting 
that neon is ionized by the
high energy emission from the central star. However, the emission may be produced by a static disk atmosphere,
a photoevaporative wind \citep{Alexander:2008, Ercolano:2010}, or a magnetically accelerated outflow \citep{Shang:2010}.
The width and the velocity shift of the [\ion{Ne}{2}] line at 12.81 $\mu$m should differ in these three cases.
The emission from a static disk atmosphere should arise at the same mean radial velocity as the central star, 
with a broadening proportional to the disk inclination (e.g. the [\ion{Ne}{2}] emission from GM Aur observed
by \citealt{Najita:2009}); whereas a photoevaporative wind -- unless the disk has a large ($\sim$30 AU) 
inner disk hole \citep{Ercolano:2010} -- will produce 
blue-shifted emission with respect to the central star ($\rm \sim 1-10~km~s^{-1}$) and, again, a broadening proportional to the disk inclination.
Blue-shifted emission, very likely originating from a photoevaporative wind, has previously been observed in 3 transitional disks  
(T Cha, CS Cha and TW Hydrae) by \cite{Pascucci:2009}. 
Finally, the emission line from a magnetically driven outflow should be blue-shifted to a larger velocity ($\sim100~km~s^{-1}$)
and much broader than emission from a disk, as observed, e.g. in Sz 73 by \cite{Pascucci:2009}.

\begin{figure*}
\includegraphics[angle=0, width=15 cm]{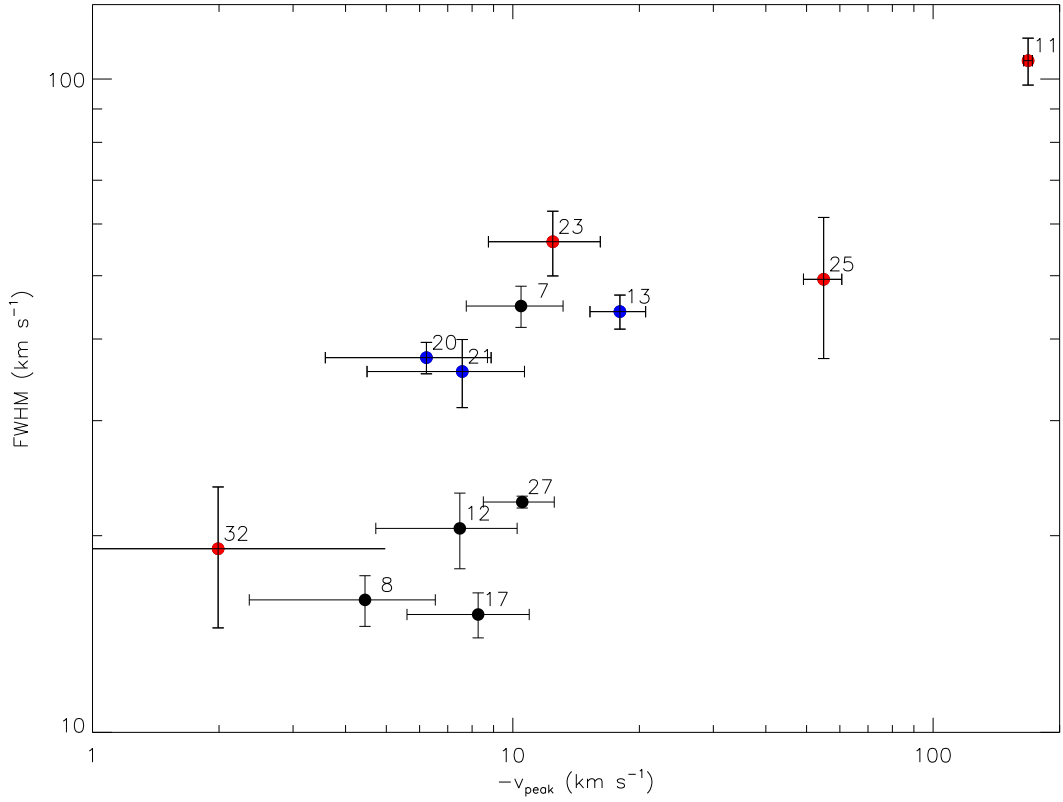}
\caption{FWHM of the [Ne II] emission line as function of the blue-shift with respect to the stellar velocity.
Colors of the symbols are as in Fig. \ref{fig:visir_vs_spitzer}. Numbers are the IDs of the stars reported in Table \ref{tab:Results}.
\label{fig:FWHM_vs_vshift}}
\end{figure*}

In Fig. \ref{fig:FWHM_vs_vshift}, we show the FWHMs of the detected lines as function of their blue-shifts with respect to the central 
star velocity. Colors represent the different infrared Classes,
as in Fig. \ref{fig:visir_vs_spitzer}. Our results rule out the possibility that the [\ion{Ne}{2}] emission is produced 
by a static disk atmosphere, since all of the detected lines are blue-shifted with respect to the central star. 
In two of the stars of our sample  (RU Lupi and IRS 60) the emission likely originates from a magnetically driven outflow, as 
demonstrated by the large blue-shift 
(v$_{peak}=168\pm$4 and v$_{peak}=55\pm 6 \rm ~km~s^{-1}$, respectively) and the large FWHM (107$\pm 9$ and 49$\pm 12 \rm~km~s^{-1}$, 
respectively) of the observed emission lines. This hypothesis is confirmed by observational evidence that indicate RU Lupi is 
the source of a well-known protostellar jet (e.g. \citealt{Takami:2001}), 
and by the proximity of IRS 60 to a Herbig-Haro objects \citep{Wu:2002}.
Furthermore, the line profile of RU Lupi (top-right panel of Fig. \ref{fig:spectra}) appears slightly asymmetric, 
with the bluer part of the profile decreasing faster than the redder part.
Similar asymmetric and strongly blue-shifted profiles have been predicted by \cite{Shang:2010}, who modeled
the [\ion{Ne}{2}] emission produced by a magnetically accelerated wind irradiated by
stellar high energy emission. Unfortunately, the signal-to-noise is too low 
to carry out a detailed comparison between our spectra and these models.
In the spectra of IRS 43 N and IRS 43 S, we also observe a secondary emission component that is 
strongly blue-shifted and asymmetric. However, although in each case the profile of this component resembles the profiles predicted 
by \cite{Shang:2010}, these highly blue-shifted and asymmetric components from both IRS 43 N and IRS 43 S are too weak 
to establish if the resemblance is significant.

\begin{figure*}
\includegraphics[angle=0, width=15 cm]{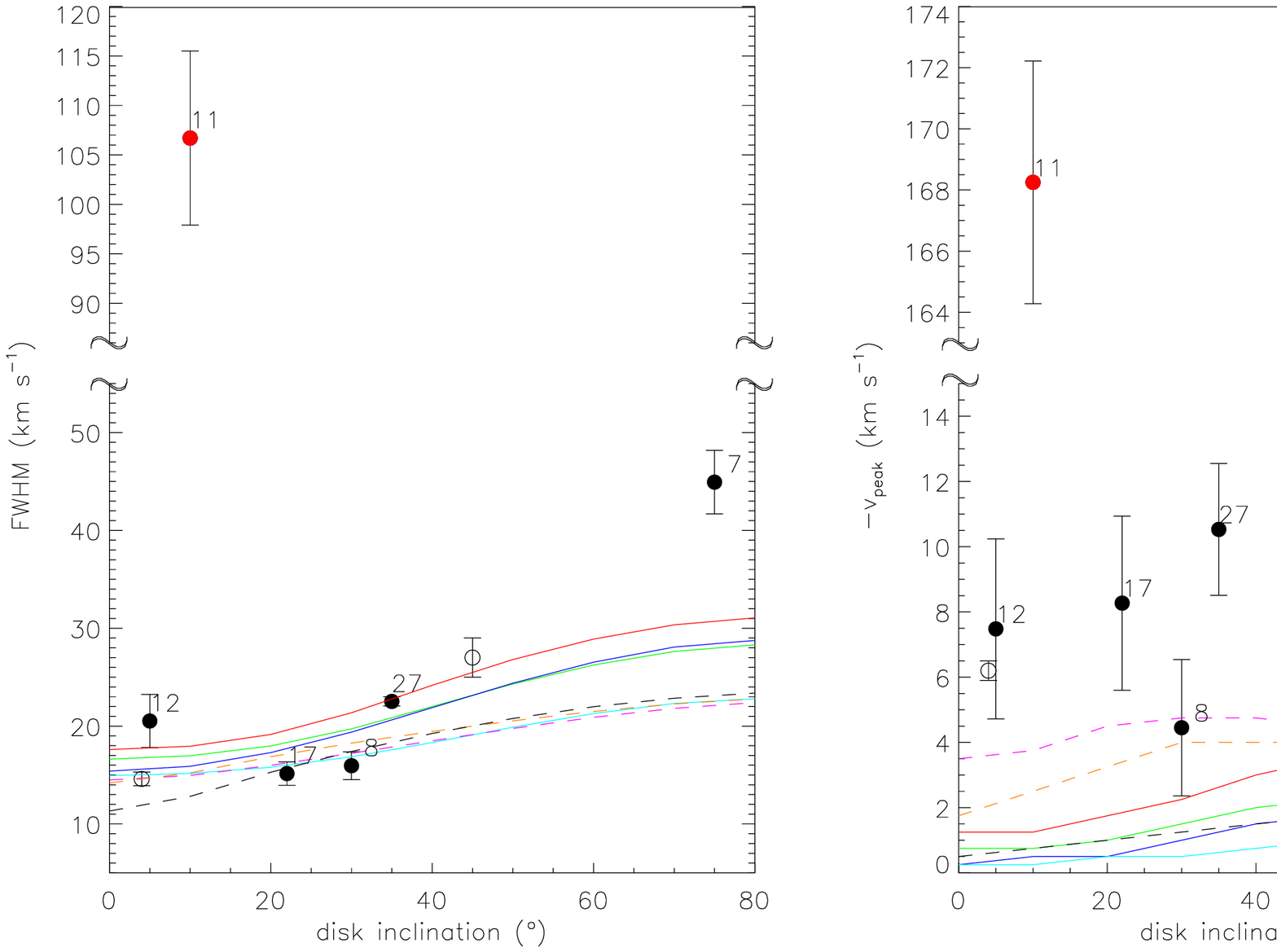}
\caption{FWHM (left panel) and blue-shift (right panel) of the [Ne II] emission as functions  of the disk inclination.
Empty symbols are from \cite{Pascucci:2009}, filled symbols from this work. Colors of the symbols are as in Fig. \ref{fig:visir_vs_spitzer}.
In both panels, we overplot 
theoretical predictions from the models of \cite{Ercolano:2010}. Specifically, red, green, blue and cyan continuous lines were computed 
for transition disks irradiated by a star with X-ray luminosity $\rm L_{X}=2\times 10^{30}~erg~s^{-1}$ and 
with an inner hole of 8.3, 14.2, 21.1 and 30.5 AUs, respectively. Magenta, orange and black dashed lines 
were computed for a primordial disk model without inner hole irradiated by a star with an X-ray luminosity of $2\times 10^{28}$, $2\times 10^{29}$ and $2\times 10^{30}~erg~s^{-1}$,
respectively. Numbers are the IDs of the stars reported in Table \ref{tab:Results} (object 11 is RU Lupi).
\label{fig:inclination}}
\end{figure*}

The emission observed from the other 10 YSOs is blue-shifted by less than 18 $\rm km~s^{-1}$.
For 5 of our targets, specifically transition and pre-transition disks, a disk origin 
is confirmed by the correlation between line FWHM and the disk inclination. The correlation 
is shown in the left panel of Fig. \ref{fig:inclination}, where we also plot two points from \cite{Pascucci:2009}. 
In this panel, we also report the relation between 
FWHM and disk inclination derived from irradiated primordial and transition disk models 
by \cite{Ercolano:2010}. Our measurements agree very 
well with the models, except in the case of RU Lupi and T Cha (the points on the top-left and top-right of the plot). 
In the first case, the emission is likely not due to a 
photoevaporative wind, but rather to a magnetically accelerated outflow, as discussed above, so the 
models of \cite{Ercolano:2010} are not applicable. In the case of T Cha the discrepancy 
between models and observational data may be related to the difference between the mass of T Cha 
(1.3 M$_{\odot}$ from \citealt{Schisano:2009})
and the mass of the central star assumed in the models (0.7 M$_{\odot}$).
Discrepancies between the models and observations are more evident in the right panel of Fig. \ref{fig:inclination}, 
where we plot the line blue-shift as a function of disk inclination. Specifically, the observed blue-shifts are higher 
than the predicted ones. Higher blue-shifts have been predicted by \cite{Alexander:2008}, who considered a warmer disk 
atmosphere (T$\sim$10,000 K) irradiated by stellar EUV radiation.

For the other 5 YSOs with an observed blue-shift of less than 18 $\rm km~s^{-1}$
(IRS 43 N, IRS 43 S, IRS 45, SSTc2dJ162145.13-234231.6 and T CrA),
the origin of the emission is less clear, because we do not know the disk inclination.
Furthermore, four of these objects are in the earliest stage of the disk evolution
(3 Class I IRS 43 N, IRS 43 S and SSTc2dJ162145.13-234231.6 and a Class II IRS 45) and the line
FWHM is larger than observed in the transition disks, so the [\ion{Ne}{2}] line at 12.81 $\mu$m could be produced by 
a jet seen with a large angle respect to the line of sight. We also note 
that in these objects errors on the blue-shift could 
be underestimated, because no radial velocities measurements are available in the literature, so we based our radial velocity
estimate on the radial velocity of the parent cloud ($\rho$ Ophiuchi) or other nearby stars (see Table \ref{obs-target}).
Observations aimed to measure accurate radial velocities and disk inclinations for these stars can help to distinguish between jet and disk 
origins.

In the left panel of Fig. \ref{fig:luminosity} we report the [\ion{Ne}{2}] luminosity as a function of the X-ray luminosity, 
with the predictions of \cite{Ercolano:2010} overplotted with a dashed line. [\ion{Ne}{2}] emission 
and X-ray luminosities do not appear correlated, as would be expected from the irradiated disk models (e.g.
\citealt{Meijerink:2008}, \citealt{Ercolano:2010}) and as has been observed by \cite{Gudel:2010}. 
However, the lack of a correlation between [\ion{Ne}{2}] emission and X-ray luminosities could be due 
to the non-homogeneity of the observed sample.
Specifically, our sample includes very different disks 
around stars of different mass that are going through different phases of the star formation process, 
while irradiated-disk models that predict a correlation focused their analysis on a single prototypical disk. 
The relation between disk properties and predicted [\ion{Ne}{2}] luminosity has been investigated by \cite{Schisano:2010}, who found that varying
disk properties can cause luminosity variations up to one order of magnitude.
Several upper limits in Fig. \ref{fig:luminosity} (left panel) lie below the [\ion{Ne}{2}] 
luminosity predicted from the model of \cite{Ercolano:2010}. 
The lack of detectable [\ion{Ne}{2}] emission in Class I and Class II sources could depend on the effects of absorption on X-ray and EUV emission.
Specifically, if EUV and X-ray emission are strongly attenuated by the optically thick inner disk layers, the amount of ionized neon 
could be less than predicted by the models. In transition disks -- like
Hen 3-600, for which the upper limit on the [\ion{Ne}{2}] luminosity is one order of magnitude below the expected value - 
the discrepancy between model and observations may be attributed to the opposite reason; namely, some of these disks have lost a large part of
their circumstellar gas, so we do not see strong emission lines in the mid-infrared. In the case of Hen 3-600, a non-detection of [\ion{Ne}{2}] 
emission is consistent both with the lack of detection of molecular CO emission \citep{Andrews:2010} and its low mass accretion rate 
(see Table \ref{obs-target}). In contrast, higher mass accretion rates and strong molecular CO emission (see \citealt{Kastner:1997,Kastner:2008,Kastner:2010, Curran:2011}) 
characterize TW Hya, V4046 Sgr and MP Mus where a strong [\ion{Ne}{2}] emission was detected.
Another possibility is that neon is ionized by EUV or soft X-ray photons mainly produced by accretion shocks, hence
the line emission a 12.81 $\mu$m strongly decreases at low mass accretion rates.

The right panel of Fig. \ref{fig:luminosity} shows a weak correlation between [\ion{Ne}{2}] luminosity and mass accretion rate, 
but this correlation is strongly influenced by just three objects (RU Lupi, IRS 43 N and IRS 43 S).
As discussed above, the [\ion{Ne}{2}] emission in RU Lupi is produced by a protostellar jet, while the origin of the emission
from IRS 43 N and IRS 43 S is uncertain. So, as suggested by \cite{Flaccomio:2009} and \cite{Gudel:2010}, the correlation between [\ion{Ne}{2}] 
luminosity and mass accretion rate may be driven by the presence of strong emission from jet sources, which are typically strong accretors.

\section{Summary and Conclusions}

We observed 32 YSOs belonging to different infrared Classes using the high resolution (R$\sim$30000) 
mid-infrared spectrograph VISIR at the VLT, with the aim of studying the origin of the [\ion{Ne}{2}] 
emission line at 12.81 $\mu$m.
We detected the line in 12 YSOs, thus tripling the number of detections of 12.81 $\mu$m
[\ion{Ne}{2}] emission in YSOs at high spectral resolution. We also collected from the literature published 
\textit{Spitzer}/IRS data for 14 of our targets and analyzed archival data for another 14 stars.
By comparing VLT/VISIR and \textit{Spitzer}/IRS data with ancillary data retrieved from the literature,
we obtained the following results:

\begin{itemize}

\item Comparison between VISIR (slit width 0.4$^{\arcsec}$) and IRS (slit width$\sim4.7^{\arcsec}$)
fluxes demonstrates that for Class I YSOs the [\ion{Ne}{2}] 
emission is mainly produced by gas located at a distance of more than 
20-40 AU (i.e., outside of the VISIR slit but within the IRS slit) from the central star
and, therefore, arises from the extended envelopes.
At these distances from the star, neon is most likely ionized by energetic shocks 
produced from protostellar outflows.

\item The same comparison (VISIR vs. IRS) shows that the [\ion{Ne}{2}] emission from transition and 
pre-transition disks is produced by gas located within a few AU from the central stars, while in 
Class II objects we observe emission both from the inner region and from the extended envelope.

\item Observed blue-shifts and FWHMs of the [\ion{Ne}{2}] line are consistent 
with emission in either a photoevaporative wind or the launching region of magnetically driven outflow. 
We associated the emission with a photoevaporative wind in 5 pre-main sequence
stars with transition and/or pre-transition disks (blue-shifts between 2 and 18 
$\rm km~s^{-1}$, FWHM between 15 and 45 $\rm km~s^{-1}$), where we observe a correlation between
line FWHM and disk inclination. Otherwise, we associated the emission to an outflow in two objects 
(blue-shifts of 55 and 168 $\rm km~s^{-1}$, FWHM of 49 and 107 $\rm km~s^{-1}$), one of which, RU Lupi, 
is a prototypical target for studies of protostellar jets. The origin of the emission in the remaining
5 objects is more uncertain. 

\item We compared the dependence of line FWHM and blue-shift on disk inclinations with the predictions of 
irradiated disk models of \cite{Ercolano:2010}.
The observed relation between line FWHM and disk inclination is consistent with the predictions of models, 
except for a single star (T Cha) whose disk is seen at high inclination angle. Line blue-shifts 
are larger than predicted from the models. 

Blue-shifted emission and a correlation of the line width with disk inclination have been also predicted 
in the case of photoevaporative wind triggered by EUV radiation \citep{Alexander:2008}.

\item We detected [\ion{Ne}{2}] emission with a blue-shifts smaller than $\rm \sim20~km~s^{-1}$ from 3 Class I 
objects and 2 Class II objects. Although these projected velocities appears small for an outflow origin, 
and would rather point toward a disk wind, we have no constraints on the inclinations of these systems and 
thus on the true velocity of the emitting material. 

\end{itemize}

Our observations demonstrate that the [\ion{Ne}{2}] emission in YSOs arises either from 
shocks formed in protostellar jets or from a photoevaporative wind in the inner disk, and 
high spatial and spectral resolutions observations are necessary to distinguish between the 
two mechanisms. Specifically, the former mechanism
prevails in early stages of the star formation process when powerful outflows are generated, while
the latter prevails in the latest stages of disk evolution. However, to understand 
if the two processes may coexist and photoevaporative winds are already active at early stages,
a better knowledge of disk properties (i.e. disk inclination, radial velocities and 
disk structure) is required.
The new instruments operating in mid-infrared and submillimeter band 
(VLT/CRIRES, Herschel, ALMA) will allow us to better study these systems and interpret our results.

For what concerns the more evolved transition and pre-transition disks, the presence of 
emission blue-shifted of $\rm \sim2-10~km~s^{-1}$ and the correlation between line width and
disk inclination are strong evidences for the photoevaporation from the disk. However,
observations in the  [\ion{Ne}{2}] line are not sufficient to understand the full physical
scenario, specifically, it is not clear if EUV emission produced by accretion shocks or
X-rays produced either by accretion shocks or coronae is the main driver of
photoevaporation and disk heating. High spectral resolution observations of other forbidden
forbidden emission lines (e.g. [\ion{O}{1}] at 6300 \AA, \citealt{Pascucci:2011}) 
emitted in the inner disk region may help to address this issue.

\begin{figure*}
\includegraphics[angle=0, width=15 cm]{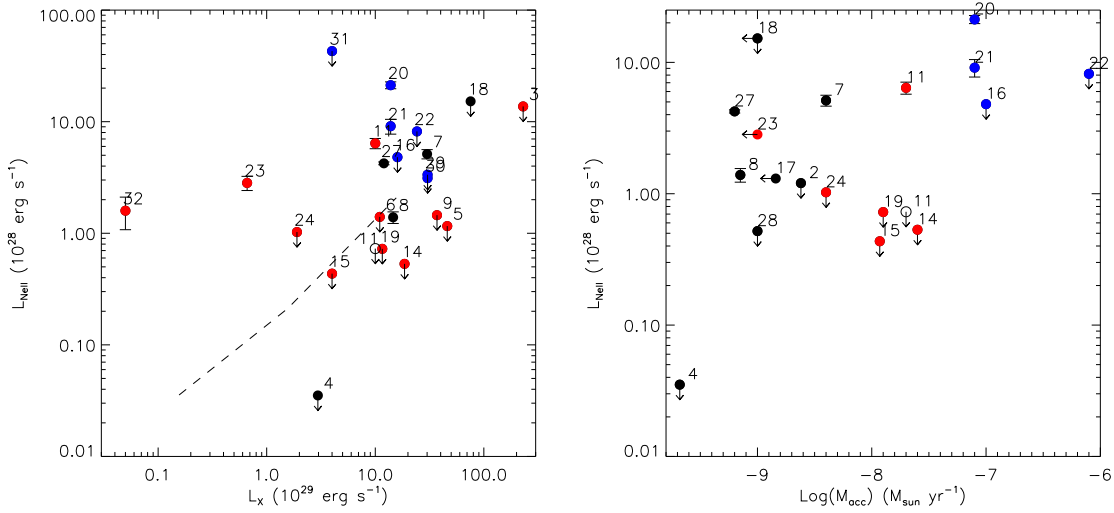}
\caption{Luminosity in the [Ne II] line as function of the X-ray luminosity in the 0.3-10 keV range
(left panel) and mass accretion rate (right panel). Colors of the symbols are as in Fig. \ref{fig:visir_vs_spitzer}, while
the empty symbol shows the upper limit for RU Lupi for a second emission component located at zero velocity.
The X-ray luminosity of the binary system IRS 43 has been assumed to be equally shared
by the two components. The dashed line on the left panels represents the relation between [\ion{Ne}{2}] and X-ray luminosity
estimated from the irradiated disk models of \cite{Ercolano:2010}. Numbers are the IDs of the stars reported in Table \ref{tab:Results}.
\label{fig:luminosity}}
\end{figure*}

Finally, we detected the [\ion{Ne}{2}] line in only 
13 YSOs out of the 32 observed. This is not surprising given that upper limits are in many 
cases of the same order of magnitude of the expected luminosity (see left 
panel of Fig. \ref{fig:luminosity}). 
Therefore, our study is biased toward stronger and 
closer sources. A higher sensitivity instrument is required to observe a larger and more complete  
sample and thereby to fully explore the space of physical parameters
of the star-disk system. The upgraded VLT/VISIR expected in the next summer will be a first step in this
direction.

\acknowledgments
This publication makes use of data products from the Two Micron All Sky Survey, 
which is a joint project of the University of Massachusetts and the Infrared Processing and 
Analysis Center/California Institute of Technology, funded by the National Aeronautics 
and Space Administration and the National Science Foundation. This work is based in part
on observations made with the Spitzer Space Telescope, obtained from the NASA/ IPAC 
Infrared Science Archive, both of which are operated by the Jet Propulsion Laboratory, 
California Institute of Technology under a contract with the National Aeronautics and 
Space Administration. We thank Dan Dicken for
his help with the data reduction of \textit{Spitzer} high resolution spectra.
This research was supported in part by NASA Astrophysics Data Analysis Program grant NNX09AC96G to RIT.
I.P. acknowledges support from NSF Astronomy \& Astrophysics research (ID: AST0908479).

{\it Facilities:} \facility{VLT:Melipal}, \facility{Spitzer}.

\begin{deluxetable}{ccccccccccl}
\tabletypesize{\scriptsize}
\rotate
\tablecaption{Observed targets.\label{obs-target}}
\tablewidth{0pt}
\tablehead{
 \colhead{Star} & \colhead{RA} & \colhead{DEC} & \colhead{region} & \colhead{d}    &  \colhead{v$_{Helio}$}       & \colhead{Class} & \colhead{inclination}
& \colhead{L$_X$\tablenotemark{h}}                     & $\rm \dot{M}_{acc}$                  &\colhead{refs} \\
 \colhead{}    & \colhead{}   &  \colhead{}   &   \colhead{}     & \colhead{(pc)} & \colhead{($\rm km~s^{-1}$)}  & \colhead{}      & \colhead{ ($^{\circ}$)} 
 &  \colhead{(10$^{29}\rm ~erg~s^{-1}$)} & \colhead{$\rm (M_{\odot}~yr^{-1}$)} & \colhead{}
 }
\startdata
         V807 Tau                  &  4:33:06.64 &  24: 9:55.0 &           Taurus & 140 &                  -                     &       Class II &  -                   &    -  &    -  &   1 -      2 -  -  -   \\
          LkCa 15                  &  4:39:17.80 &  22:21:03.5 &           Taurus & 140 &        17.0$\pm$1.2                    &            pTr &  42\tablenotemark{e} &    -  &  -8.6 &   1  3     4  4 -   4  \\
  IRAS 08267-3336                  &  8:28:40.70 & -33:46:22.2 &       Gum Nebula & 450 &                  -                     &       Class II &  -                   & 230.0 &    -  &   5 -      5 -   6 -   \\
      Hen 3-600 A                  & 11:10:27.88 & -37:31:52.0 &           TW Hya &  39 &        15.6$\pm$0.2                    &             Tr &  36\tablenotemark{e} &   3.0 &  -9.7 &   7  8     9 10 11 12  \\
           WX Cha                  & 11:09:58.74 & -77:37:08.9 &            Cha I & 178 &                  -                     &       Class II &  -                   &  46.0 &    -  &   6 -     13 -   6 -   \\
           XX Cha                  & 11:11:39.66 & -76:20:15.3 &            Cha I & 178 &                  -                     &       Class II &  -                   &  11.0 &    -  &   6 -     13 -   6 -   \\
            T Cha                  & 11:57:13.49 & -79:21:31.4 &   $\epsilon$ Cha & 109 &        14.0$\pm$1.3                    &             Tr &  75\tablenotemark{f} &  30.0 &  -8.4 &   8 14    15 16  6 16  \\
           MP Mus                  & 13:22:07.53 & -69:38:12.2 &   $\epsilon$ Cha & 103 &        11.6$\pm$0.2                    &            pTr &  30\tablenotemark{g} &  14.6 &  -9.1 &   8  8    17 18 19 12  \\
          IM Lupi                  & 15:56:09.22 & -37:56:05.8 &          Lupus 2 & 150 &                  -                     &       Class II &  50\tablenotemark{g} &  37.0 &    -  &  20 -     21 21  6 -   \\
            SZ 73                  & 15:47:56.94 & -35:14:34.7 &          Lupus 1 & 150 &        -3.3$\pm$2.5                    &       Class II &  -                   &    -  &    -  &  20 22    23 -  -  -   \\
          RU Lupi                  & 15:56:42.30 & -37:49:15.4 &          Lupus 2 & 150 &        -1.9$\pm$0.2                    &    ClassII/Jet &  10\tablenotemark{h} &  10.0 &  -7.7 &  20 24 25/26 24  6 12  \\
  RX J1615.3-3255                  & 16:15:20.23 & -32:55:05.1 &            Lupus & 150 &        -2.4$\pm$1.0                    &             Tr &   5\tablenotemark{f} &    -  &    -  &  20 27    28 28 -  -   \\
  SSTc2dJ162145.1\tablenotemark{a} & 16:21:45.13 & -23:42:31.6 &       $\rho$ Oph & 120 &        -6.3$\pm$1.7\tablenotemark{b,c} &       Class I  &  -                   &    -  &    -  &  29 30    36 -  -  -   \\
  DoAr 25/GY92 17                  & 16:26:23.68 & -24:43:13.9 &       $\rho$ Oph & 120 &        -6.3$\pm$1.7\tablenotemark{b,c} &       Class II &  59\tablenotemark{e} &  18.6 &  -7.6 &  29 30    31 32 31 31  \\
   WL 10/GY92 211                  & 16:27:09.11 & -24:34:08.1 &       $\rho$ Oph & 120 &        -6.3$\pm$1.7\tablenotemark{b,c} &       Class II &  -                   &   4.0 &  -7.9 &  29 30    31 -  31 31  \\
Elias 29/GY92 214                  & 16:27:09.43 & -24:37:18.7 &       $\rho$ Oph & 120 &        -6.3$\pm$1.7\tablenotemark{b,c} &        Class I &  -                   &  16.0 &  -7.0 &  29 30    31 -  31 31  \\
            SR 21                  & 16:27:10.28 & -24:19:12.7 &       $\rho$ Oph & 120 &        -6.3$\pm$1.7\tablenotemark{b,c} &             Tr &  22\tablenotemark{e} &    -  &   8.8 &  29 30    33 32 -  34  \\
            WL 19                  & 16:27:11.71 & -24:38:32.1 &       $\rho$ Oph & 120 &        -6.3$\pm$1.7\tablenotemark{b,c} &            pTr &  -                   &  75.5 &   9.0 &  29 30    31 -  31 31  \\
   WL 20/GY92 240                  & 16:27:15.88 & -24:38:43.4 &       $\rho$ Oph & 120 &        -6.3$\pm$1.7\tablenotemark{b,c} &       Class II &  -                   &  11.6 &  -7.9 &  29 30    31 -  31 31  \\
  IRS 43/GY92 265                  & 16:27:26.94 & -24:40:50.8 &       $\rho$ Oph & 120 &        -7.2$\pm$1.5\tablenotemark{d}   &        Class I &  -                   &  27.7 &  -7.1 &  29 35    31 -  31 31  \\
  IRS 44/GY92 269                  & 16:27:28.03 & -24:39:33.5 &       $\rho$ Oph & 120 &        -5.8$\pm$1.5\tablenotemark{d}   &        Class I &  -                   &  24.2 &  -6.1 &  29 35    31 -  31 31  \\
  IRS 45/GY92 273                  & 16:27:28.44 & -24:27:21.0 &       $\rho$ Oph & 120 &        -3.1$\pm$1.5\tablenotemark{b,c} &       Class II &  -                   &   0.7 &   9.0 &  29 35    31 -  31 31  \\
  IRS 47/GY92 279                  & 16:27:30.18 & -24:27:43.4 &       $\rho$ Oph & 120 &        -3.1$\pm$1.5\tablenotemark{d}   &       Class II &  -                   &   1.9 &  -8.4 &  29 35    31 -  31 31  \\
   IRS 60/WSB 71a                  & 16:31:30.88 & -24:24:40.0 &       $\rho$ Oph & 120 &        -6.3$\pm$1.7\tablenotemark{b,c} &   Class II/Jet &  -                   &    -  &    -  &  29 31 36/37 -  -  -   \\
          DoAr 44                  & 16:31:33.46 & -24:27:37.3 &       $\rho$ Oph & 120 &        -6.3$\pm$1.7\tablenotemark{b,c} &            pTr &  45\tablenotemark{e} &    -  &    -  &  29 30 38/32 32 -  -   \\
        V4046 Sgr                  & 18:14:10.48 & -32:47:34.4 &      $\beta$ Pic &  73 &        -6.9$\pm$0.2                    &             Tr &  35\tablenotemark{e} &  12.0 &  -9.2 &  39 40    41 42 43 12  \\
  RX J1842.9-3532                  & 18:42:57.98 & -35:32:42.7 &              CrA & 130 &         1.2$\pm$1.0                    &             Tr &  -                   &    -  &  -9.0 &  44 44    45 -  -   6  \\
         CrA IRS5                  & 19:01:48.02 & -36:57:22.4 &              CrA & 130 &         1.0$\pm$1.0                    &    Class I/Jet &  -                   &  30.3 &    -  &  46 46 47/48 - 4       \\
         R CrA 7B                  & 19:01:56.39 & -36:57:28.4 &              CrA & 130 &         1.0$\pm$1.0                    &      Class 0/I &  -                   &   4.0 &    -  &  46 46    47 -  47 -   \\
            T CrA                  & 19:01:58.78 & -36:57:49.9 &              CrA & 130 &         1.0$\pm$1.0                    &   Class II/Jet &  -                   &   0.1 &    -  &  46 46 47/48 -  47 -   \\
\enddata
\tablecomments{ Stellar coordinates are retreived from the 2MASS catalog \citep{Skrutskie:2006}, except
for CrA IRS5 and R CrA 7B \citep{Forbrich:2007}. References:
(1) \cite{Kenyon:1994};  (2) \cite{Furlan:2009}; (3) \cite{Hartmann:1987};  (4) \cite{Espaillat:2007a}; (5) \cite{Pettersson:2008};
(6) \cite{Gudel:2010}; (7) \cite{Kastner:1997};(8) \cite{Torres:2006};(9) \cite{Uchida:2004};(10) \cite{Andrews:2010};
(11) \cite{Huenemoerder:2007}; (12) \cite{Curran:2011}; (13) \cite{Manoj:2011}; (14) \cite{Guenther:2007}; 
(15) \cite{Brown:2007}; (16) \cite{Schisano:2009}; (17) \cite{Cortes:2009}; (18) \cite{Kastner:2010}; (19) \cite{Argiroffi:2009};
(20) \cite{Comeron:2008}; (21) \cite{Pinte:2008}; (22)\cite{Melo:2003}; (23)\cite{Hughes:1994}; (24) \cite{Stempels:2002}; (25) \cite{Schegerer:2009}; (26)\cite{Takami:2001}
(27) \cite{Wichmann:1999}; (28) \cite{Merin:2010};(29) \cite{Lombardi:2008};(30) \cite{Doppmann:2003};(31) \cite{Flaccomio:2009}; 
(32) \cite{Andrews:2009};(33) \cite{Eisner:2009}; (34) \cite{Natta:2006};
(35) \cite{Covey:2006};(36) \cite{Evans:2009};(37) \cite{Wu:2002};(38) \cite{Espaillat:2010};
(39) \cite{Torres:2008}; (40) \cite{Quast:2000}; (41) \cite{Schutz:2009};(42) \cite{Kastner:2008};(43) \cite{Argiroffi:2011};
(44) \cite{Neuhauser:2000};(45) \cite{Neuhauser:2008}; (46) \cite{Hughes:2010}; (47) \cite{Forbrich:2007}; (48) \cite{Wang:2004a};
}
\tablenotetext{a}{Full stellar name SSTc2dJ162145.13-234231.6.}
\tablenotetext{b}{Mean radial velocity of ten memebers of the $\rho$ Oph star forming region measured by \cite{Doppmann:2003}.}
\tablenotetext{c}{The heliocentric velocity has been derived from the velocity with respect to the local standard of rest using the IRAF
task rvcorrect.}
\tablenotetext{d}{Assumed equal to the radial velocity of IRS 47.}
\tablenotetext{e}{Measured by submillimeter interferometric observatins of the disk.}
\tablenotetext{f}{Derived from the fit of the SED.}
\tablenotetext{g}{Measured by scattered-light imaging of the disk.}
\tablenotetext{h}{Assumed equal to the inclination of the stellar rotation axes.}
\tablenotetext{i}{Unabsorbed X-ray luminosity in the range 0.3-10 KeV.}
\end{deluxetable}

\begin{deluxetable}{lcccccccc}
\tabletypesize{\scriptsize}
\tablecaption{Observation log. \label{Obs-log}}
\tablewidth{0pt}
\tablehead{
\colhead{Star} & \colhead{day} & \colhead{U.T.} & \colhead{airmass} & \colhead{t$_{exp}$} & \colhead{standard} & \colhead{U.T.} & \colhead{airmass} & \colhead{t$_{exp}$} \\
 \colhead{}    & \colhead{(yy-mm-dd)} & \colhead{(hh:mm)} & \colhead{}   &\colhead{(s)} & \colhead{} & \colhead{(hh:mm)} & \colhead{(s)}  \\
}
\startdata
T Cha            & 2009-06-02 & 00:01       & 1.7-1.8  & 1800 & HD 92305 & 23:05 &1.7& 360 \\
IRS 43 S         & 2009-06-02 & 01:25       & 1.2-1.3  & 450  & HD 149447 & 01:02/02:42 & 1.4/1.1 &360$\times2$ \\
IRS 43 N         & 2009-06-02 & 01:58       & 1.1-1.2  & 450  & HD 149447 & 02:42 & 1.1 &360 \\
DoAr 25          & 2009-06-02 & 03:09       & 1.0-1.05 & 1700 & HD 151680 & 04:35 & 1.02 &360  \\  
DoAr 25          & 2009-06-02 & 04:44       & 1.0-1.3 & 5500  & HD 151680 & 04:35/07:32 & 1.0/1.2-1.3 &360$\times2$ \\  
IRS 44           & 2009-06-02 & 07:50       & 1.4-1.8 & 1800 &  HD 151680 & 07:32/08:50 & 1.2-1.3/1.6-1.7 & 360$\times2$  \\ 
CrA IRS 5        & 2009-06-02 & 09:38       & 1.2-1.4 & 1800 &  HD 177716 & 09:09 & 1.1 & 360  \\ 
Sz 73            & 2009-06-02 & 23:46        & 1.4-1.6 & 1800 &  HD 139127 & 23:20 & 1.65-1.71 &360  \\ 
WL 10            & 2009-06-03 & 01:14        & 1.0-1.4 & 3600 & HD 151680 & 00:41/03:53 & 1.6/1.0 &360$\times2$  \\ 
Kalliope         & 2009-06-03 & 03:35        & 1.0-1.1 & 360  &   -       &  -          &  -       \\
RX J1842.9-3532  & 2009-06-03 & 04:19        & 1.1-1.2 & 1800 & HD 151680 & 03:53 & 1.0 &360  \\ 
WL 10            & 2009-06-03 & 05:56        & 1.0-1.4 & 3600 & HD 151680 & 07:49 & 1.3 &360  \\ 
WL 20            & 2009-06-03 & 08:08        & 1.6-2.0 & 1800 &  HD 151680 & 09:02 & 1.8& 360 \\ 
RX J1842.9-3532  & 2009-06-03 & 09:34        & 1.3-1.7 & 1800 & HD 151680 & 09:02 & 1.8& 360 \\ 
Titan            & 2009-06-03 & 22:43       &    1.2     & 300  &     -       &  -          &  -       \\
IRS 47           & 2009-06-03 & 00:36       & 1.3-1.6 & 1800 &  HD 151680 & 01:35 & 1.3 & 360 \\ 
IRS 47           & 2009-06-04 & 02:01       & 1.0-1.2 & 1800 &  HD 151680 & 01:35/03:50 & 1.3/1.0 & 360$\times2$   \\ 
IRS 45           & 2009-06-04 & 04:48       & 1.0-1.1 & 3600 &  HD 151680 & 03:50/06:32 & 1.0/1.1 & 360$\times2$  \\ 
IRS 45           & 2009-06-04 & 06:52       & 1.2-1.7 & 3600 &  HD 151680 & 06:32/08:31 & 1.1/1.6 & 360  \\ 
V4046 Sgr        & 2009-06-04 & 08:55       & 1.3-1.6 & 2700  & HD 169916 & 10:12       & 1.6-1.8 & 600 \\ 
IRAS 08267-3336  & 2009-12-15 & 07:18       & 1.01-1.02& 1980 & HD 70555  & 08:45        & 1.03-1.06 & 180 \\  
IRAS 08267-3336  & 2009-12-15 & 06:12       & 1.02-1.07& 1980 & HD 70555 & 05:48         & 1.1     & 180\\
XX Cha           & 2010-01-04 & 06:34       &   1.7    & 1980 & HD 92305 & 05:50         & 1.8     & 180 \\
WX Cha           & 2010-01-08 & 04:48       &   1.7  &   1980 & HD 92305 & 05:57         & 1.7     &180  \\
WX Cha           & 2010-01-08 & 06:44       & 1.8-2.0 &  1980 & HD 92305 & 07:48         & 1.8     &180  \\
XX Cha           & 2010-01-09 & 03:58       & 1.9-2.1  & 1980 & HD 92305 & 05:02         & 1.9     & 180 \\
IM Lupi          & 2010-01-30 & 07:48       & 1.4-1.7  & 1980 & HD 139127 & 08:51         & 1.3     & 180 \\
IM Lupi          & 2010-01-31 & 07:54       & 1.4-1.6  & 1980 & HD 139127 & 07:20         & 1.7-1.8   & 180 \\
MP Mus           & 2010-06-03 & 23:18       &1.4-1.5 & 3600  & HD 123139 & 22:53 & 1.3 & 360 \\
IRS 44           & 2010-06-04 & 01:05       &1.2-1.4 & 1800 & HD 151680 & 02:04 & 1.2 &360 \\
Elias 29         & 2010-06-04 & 02:29       &1.0-1.1 & 3600 &  HD 151680 & 04:10 & 1.0 & 720 \\ 
DoAr 44          & 2010-06-04 & 05:20       & 1.0-1.2 & 3600 &  HD 151680 & 04:10/07:02 & 1.0/1.2 &720/360 \\
RU Lupi          & 2010-06-04 & 07:37       & 1.5-1.9 & 3600 & HD 150798 & 08:35 & 1.8 & 180 \\
T CrA            & 2010-06-04 & 09:22       & 1.2-1.6 & 7200 & HD 169916 & 8:52 & 1.2-1.3 & 600 \\
Hen 3-600 A      & 2010-06-04 & 23:10       &1.0-1.1  & 3600 & HD 93813 & 22:48 & 1.0 & 360 \\
SR 21            & 2010-06-05 & 01:36       &1.0-1.2  & 3600 & HD 151680 & 03:18 & 1.2 &360 \\
RU Lupi          & 2010-06-05 & 03:44       & 1.0-1.1 & 3600 & HD 151680 & 03:18/05:24 & 1.2/1.0 &360/720 \\
SR 21            & 2010-06-05 & 05:49       &1.0-1.3  & 3600 & HD 151680 & 05:24/07:30 & 1.0/1.1-1.2& 720/360 \\
R CrA 7B         & 2010-06-05 & 08:39       & 1.1-1.4 & 3600 & HD 169916 & 08:08 & 1.1 & 360 \\
RXJ1615.3-3255   & 2010-06-18 & 01:55       & 1.0     & 1620 & HD 149447 & 03:14 & 1.0 & 180 \\
SSTc2dJ162145.1-234232 & 2010-07-10 & 03:23 & 1.1-1.2 & 1620 & HD 139163 & 04:18 & 1.4 & 360 \\
IRS 60           & 2010-07-10 & 4:58        & 1.3-1.6 & 1980 & HD 151680 & 05:59 & 1.5 & 180 \\
WL 19            & 2010-07-11 & 03:12       & 1.05-1.12 & 1620 & HD 151680 & 04:08 & 1.1 & 180 \\
WL 19            & 2010-07-11 & 04:56       & 1.3-1.6 & 1620 & HD 151680 & 05:48 & 1.5 & 180 \\
WL 20            & 2010-07-12 & 02:25       & 1.0-1.1 & 1980 & HD 151680 & 03:27 & 1.1 & 180 \\
V807 Tau         & 2010-12-25 & 02:52       &   1.5   & 1980 & HD 28305   & 03:50 & 1.4 &180 \\
LkCa 15          & 2011-01-12 & 02:27       &   1.5   & 1620 & HD 28305   & 01:52 & 1.4 & 180 \\
LkCa 15          & 2011-01-12 & 03:30       & 1.6-1.8 & 1620 & HD 28305   & 04:24 & 1.9 & 180 \\
\enddata
\end{deluxetable}

\begin{deluxetable}{ccccccccc}
\tabletypesize{\scriptsize}
\tablecaption{Results.\label{tab:Results}}
\tablewidth{0pt}
\tablehead{
\colhead{ID} & \colhead{Star} & \colhead{Flux}     & \colhead{FWHM}  &  \colhead{v$_{peak}$}  &  \colhead{A$_J$} &\colhead{L$_{NeII}$} & \colhead{Flux (Spitzer)}  & \colhead{refs} \\
 \colhead{} & \colhead{}    & \colhead{$10^{-14}~erg~cm^{-2}~s^{-1}$} &  \colhead{$km~s^{-1}$} & \colhead{$km~s^{-1}$}   & \colhead{mag} &\colhead{$10^{28}~erg~s^{-1}$} & \colhead{$10^{-14}~erg~cm^{-2}~s^{-1}$}& \\
}
\startdata
 1 &                  V807 Tau &                                        $<$1.1 &                                            -  &                                            -  &      0.2 &      $<$ 2.6 &                -  &   1 -  \\
 2 &                   LkCa 15 &                                        $<$0.5 &                                            -  &                                            -  &      0.3 &      $<$ 1.2 &    0.28$\pm$ 0.02 &   1 11 \\
 3 &           IRAS 08267-3336 &                                        $<$0.5 &                                            -  &                                            -  &      1.4 &      $<$13.8 &    1.40$\pm$ 0.08 &   4 11 \\
 4 &               Hen 3-600 A &                                        $<$0.2 &                                            -  &                                            -  &      0.2 &      $<$0.03 &           $< 0.5$ &   2 11 \\
 5 &                    WX Cha &                                        $<$0.2 &                                            -  &                                            -  &      0.6 &      $<$0.03 &    0.57$\pm$ 0.06 &   1 11 \\
 6 &                    XX Cha &                                        $<$0.2 &                                            -  &                                            -  &      0.3 &      $<$0.03 &    0.48$\pm$ 0.02 &   1 11 \\
 7 &                     T Cha &                                   3.4$\pm$0.3 &                                $ 44.9\pm 3.2$ &                                 -10.5$\pm$2.7 &      0.8 &  5.1$\pm$0.5 &    3.20$\pm$ 0.21 &   5 12 \\
 7 &    T Cha\tablenotemark{a} &                                   3.1$\pm$0.2 &                                  39.4$\pm$1.9 &                                  -4.7$\pm$2.5 &      0.8 &  4.7$\pm$0.3 &    3.20$\pm$ 0.21 &   5 12 \\
 8 &                    MP Mus &                                   1.1$\pm$0.1 &                                $ 15.9\pm 1.4$ &                                  -4.4$\pm$2.1 &      0.1 &  1.4$\pm$0.2 &           $< 2.0$ &   2 13 \\
 9 &                   IM Lupi &                                        $<$0.5 &                                            -  &                                            -  &      0.2 &      $<$ 1.4 &    1.07$\pm$ 0.05 &   4 11 \\
10 &                     SZ 73 &                                        $<$0.6 &                                            -  &                                            -  &      0.8 &      $<$ 1.8 &    1.60$\pm$ 0.24 &   6 12 \\
10 &    Sz 73\tablenotemark{a} &                                   1.9$\pm$0.2 &                                  72.8$\pm$6.4 &                                     -97$\pm$4 &      0.8 &  5.4$\pm$0.6 &    1.60$\pm$ 0.24 &   6 12 \\
11 &                   RU Lupi &          2.4$\pm$0.2($<$0.3)\tablenotemark{b} &                                $106.7\pm 8.8$ &                                -168.2$\pm$4.0 &      0.0 &  6.4$\pm$0.7 &    2.60$\pm$ 0.83 &   2  4 \\
12 &           RX J1615.3-3255 &                                   1.4$\pm$0.2 &                                $ 20.5\pm 2.7$ &                                  -7.5$\pm$2.8 &      0.3 &  3.8$\pm$0.7 &    2.76$\pm$ 0.46 &   7 11 \\
13 &           SSTc2dJ162145.1 &                                   5.8$\pm$0.5 &                                $ 44.1\pm 2.6$ &                                 -18.0$\pm$2.7 &      2.8 & 15.0$\pm$1.2 &   11.83$\pm$ 0.46 &   8 11 \\
14 &           DoAr 25/GY92 17 &                                        $<$0.3 &                                            -  &                                            -  &      0.7 &      $<$ 0.5 &    0.50$\pm$ 0.10 &   3  3 \\
15 &            WL 10/GY92 211 &                                        $<$0.1 &                                            -  &                                            -  &      4.5 &      $<$ 0.4 &    1.60$\pm$ 0.30 &   3  3 \\
16 &         Elias 29/GY92 214 &                                        $<$0.5 &                                            -  &                                            -  &     11.4 &      $<$ 4.8 &           $< 8.5$ &   3  3 \\
17 &                     SR 21 &                                   0.5$\pm$0.1 &                                $ 15.1\pm 1.2$ &                                  -8.3$\pm$2.7 &      2.3 &  1.3$\pm$0.1 &           $< 3.0$ &   1  4 \\
18 &                     WL 19 &                                        $<$0.8 &                                            -  &                                            -  &     16.3 &      $<$15.2 &           $< 0.8$ &   3  3 \\
19 &            WL 20/GY92 240 &                                        $<$0.2 &                                            -  &                                            -  &      4.1 &      $<$ 0.7 &    6.28$\pm$ 0.25 &   3  3 \\
20 &         IRS 43/GY92 265 S &                                   3.7$\pm$0.3 &                                $ 37.5\pm 2.1$ &                                  -6.2$\pm$2.7 &      8.2 & 21.2$\pm$1.5 &   45.30$\pm$ 1.70 &   3  3 \\
21 &         IRS 43/GY92 265 N &                                   1.6$\pm$0.2 &                                $ 35.7\pm 4.3$ &                                  -7.6$\pm$3.1 &      8.2 &  9.1$\pm$1.4 &   45.30$\pm$ 1.70 &   3  3 \\
22 &           IRS 44/GY92 269 &                                        $<$0.7 &                                            -  &                                            -  &     12.8 &      $<$ 8.2 &    8.00$\pm$ 2.40 &   3  3 \\
23 &           IRS 45/GY92 273 &                                   0.6$\pm$0.1 &                                $ 56.4\pm 6.4$ &                                 -12.4$\pm$3.7 &      6.6 &  2.9$\pm$0.4 &    2.19$\pm$ 0.41 &   3  3 \\
24 &           IRS 47/GY92 279 &                                        $<$0.2 &                                            -  &                                            -  &      7.4 &      $<$ 1.0 &    2.01$\pm$ 0.51 &   3  3 \\
25 &            IRS 60/WSB 71a &          0.9$\pm$0.3($<$0.2)\tablenotemark{b} &                                $ 49.4\pm12.0$ &                                 -54.9$\pm$5.7 &      2.5 &  2.3$\pm$0.8 &    1.91$\pm$ 0.09 &   1 11 \\
26 &                   DoAr 44 &                                        $<$0.3 &                                            -  &                                            -  &      1.2 &      $<$ 0.5 &                -  &   8 -  \\
27 &                 V4046 Sgr &                                   6.6$\pm$0.2 &                                $ 22.5\pm 0.5$ &                                 -10.5$\pm$2.0 &      0.0 &  4.2$\pm$0.1 &    7.68$\pm$ 0.22 &   2 11 \\
28 &           RX J1842.9-3532 &                                        $<$0.2 &                                            -  &                                            -  &      0.3 &      $<$ 0.5 &    0.43$\pm$ 0.13 &   9 13 \\
29 &                CrA IRS5 A &                                        $<$0.6 &                                            -  &                                            -  &      7.0 &      $<$ 3.3 &   20.32$\pm$ 0.65 &  10 11 \\
30 &                CrA IRS5 B &                                        $<$0.6 &                                            -  &                                            -  &      7.0 &      $<$ 3.1 &   20.32$\pm$ 0.65 &  10 11 \\
31 &                  R CrA 7B &                                        $<$0.5 &                                            -  &                                            -  &     26.0 &      $<$42.9 &   10.51$\pm$ 0.42 &  10 11 \\
32 &                     T CrA &                                   0.7$\pm$0.2 &                                $ 19.1\pm 4.7$ &                                  -2.0$\pm$3.0 &      0.7 &  1.6$\pm$0.5 &    8.88$\pm$ 1.30 &  10 11 \\
\enddata
\tablecomments{For the Ne II detections, fluxes, FWHMs and the line peak velocity (v$_{peak}$) are
calculated by fitting the spectrum with a gaussin profile for the line and a first order polynomial for the continuum.
The line peak velocity is measured with respect to a reference system comoving with the star. Errors (1$\sigma$)
are derived from the best-fit procedure. The Ne II luminosity are corrected for the extinction as discussed in the text.
The last column reports the references for the A$_J$ and the Ne II flux as measured
from Spitzer low and medium resolution spectoscopy: 
(1) \cite{Furlan:2009}; (2) \cite{Curran:2011}; (3) \cite{Flaccomio:2009}; (4) \cite{Gudel:2010}; (5) \cite{Schisano:2009};
(6) \cite{Hughes:1994}; (7) \cite{Merin:2010}; (8) \cite{Evans:2009}; (9) \cite{Carpenter:2008}; (10) \cite{Forbrich:2007};
(11) this work; (12) \cite{Lahuis:2007}; (13) \cite{Pascucci:2007}}
\tablenotetext{a}{Results obtained from a new reduction and calibration of the observations carried out by \cite{Pascucci:2009}}
\tablenotetext{b}{Upper limit for a disk emission centered at 0 $km~s^{-1}$.}
\end{deluxetable}

\end{document}